\documentclass{aa}

\usepackage{graphicx}
\usepackage{txfonts}
\usepackage{lipsum}
\usepackage{subcaption}         
\usepackage{lscape}             
\usepackage{placeins}           
\usepackage{xcolor}

\def\chisqr{\hbox{$\chi^2_{\rm r}$}}

\def\me{\hbox{${\rm M}_{\oplus}$}}
\def\re{\hbox{${\rm R}_{\oplus}$}}
\def\mspy{\hbox{${\rm M}_{\odot}$\,yr$^{-1}$}}
\def\msw{\hbox{$\dot {\rm M}_{\odot}$}}

\def\mstar{\hbox{$M_{\star}$}}
\def\rstar{\hbox{$R_{\star}$}}

\def\teff{\hbox{$T_{\rm eff}$}}
\def\logg{\hbox{$\log g$}}

\def\vD{\hbox{$v_{\rm D}$}}

\def\ms{\hbox{m\,s$^{-1}$}}

\def\kms{\hbox{km\,s$^{-1}$}}
\def\gpcc{\hbox{g\,cm$^{-3}$}}
\def\vsini{\hbox{$v \sin i$}}

\def\mic{\hbox{$\mu$m}}

\def\emr{}
\def\emn{}
\def\Bl{\hbox{$B_{\rm \ell}$}}
\def\Bd{\hbox{$B_{\rm d}$}}

\def\degr{\hbox{$^\circ$}}

\def\Prot{\hbox{$P_{\rm rot}$}}

\newcommand{\hei}{\hbox{He$\;${\sc i}}}

\newcommand{\pab}{\hbox{Pa${\beta}$}}

\begin{document}

\title{Six-yr SPIRou monitoring of the young planet-host dwarf AU~Mic} 

   \author{J.-F.~Donati\inst{1} 
      \and P.I.~Cristofari\inst{2}
      \and C.~Moutou\inst{1}
      \and A.~L'Heureux\inst{3}
      \and N.J.~Cook\inst{3}
      \and E.~Artigau\inst{3}
      \and S.H.P.~Alencar\inst{4}
      \and E.~Gaidos\inst{5,6}
      \and A.~Vidotto\inst{2}
      \and P.~Petit\inst{1}
      \and A.~Carmona\inst{1}
      \and T.~Ray\inst{7}
      \and the SPIRou science team
          } 
   \institute{Univ.\ de Toulouse, CNRS, IRAP, 14 avenue Belin, 31400 Toulouse, France\\ 
                       \email{jean-francois.donati@irap.omp.eu} 
         \and Leiden Observatory, Leiden University, Niels Bohrweg 2, 2333 CA Leiden, the Netherlands 
         \and Universit\'e de Montr\'eal, D\'epartement de Physique, IREX, Montr\'eal, QC H3C 3J7, Canada 
         \and Departamento de F\'{\i}sica -- ICEx -- UFMG, Av. Ant\^onio Carlos, 6627, 30270-901 Belo Horizonte, MG, Brazil
         \and Department of Earth Sciences, University of Hawai'i at M\={a}noa, Honolulu, Hawai'i 96822 USA 
         \and Institute for Astrophysics, University of Vienna, 1180 Vienna, Austria 
         \and Dublin Institute for Advanced Studies, Astronomy \& Astrophysics Section, 31 Fitzwilliam Place, Dublin D02 XF86, Ireland 
             } 

\date{Submitted 2025 May 02 -- Accepted 2025 July 01 } 

\abstract{ 
In this paper we revisit our spectropolarimetric and velocimetric analysis of the young M dwarf AU~Mic based on data collected with SPIRou at the Canada-France-Hawaii telescope, 
over a monitoring period of 2041~d from 2019 to 2024.  The longitudinal magnetic field, the small-scale magnetic field, and the differential temperature 
of AU~Mic, derived from the unpolarized and circularly-polarized spectra, were clearly modulated with the stellar rotation period, with a pattern that evolved over time.  
The magnetic modeling with Zeeman-Doppler imaging provides a consistent description of the global field of AU~Mic that agrees not only with the Least-Squares Deconvolved profiles 
of the circularly-polarized and unpolarized spectral lines, but also with the small-scale field measurements derived from the broadening of spectral lines, for each of the 11 subsets of the 
full data.  We find that the large-scale field was mostly poloidal, with a dominant dipole component slightly tilted to the rotation axis which decreased from 1.4 to 1.1~kG before increasing 
at the end of the campaign.  The average small-scale field followed a similar trend, decreasing from 2.8 to 2.6~kG then rising.  {\emr The long-term magnetic evolution we report for AU~Mic 
suggests that, if cyclic, the cycle period is significantly longer than 6~yr.} 
From velocimetric data, we derived improved mass estimates for the two transiting planets, respectively equal to $M_b=6.3^{+2.5}_{-1.8}$~\me\ and $M_c=11.6^{+3.3}_{-2.7}$~\me, yielding 
very contrasting densities of $0.32^{+0.13}_{-0.10}$ and $2.9^{+1.1}_{-0.8}$~\gpcc, and a new 90\% confidence upper limit of 4.9~\me\ for candidate planet d (period 12.7~d) suspected to 
induce the transit-timing variations of b and c.  We also confirm our claim regarding candidate planet e orbiting with a period of $33.11\pm0.06$~d, albeit with a smaller mass of 
$M_e=21.1^{+5.4}_{-4.3}$~\me.  
} 


\maketitle



\section{Introduction}
\label{sec:int}

Comparing observations of young exoplanets with those of older counterparts provides constraints on theoretical models of planet formation and evolution.  However, very young stars 
are often active and rapidly rotating, rendering their close-in planets challenging to detect and characterize through the usual photometric technique from dedicated 
spacecraft (for planets transiting in front of their host stars) and with the velocimetric method involving ground based ultra-stable high-resolution spectrometers.  For this reason, 
only a handful of planets around stars younger than 20~Myr have been reliably confirmed so far, implying that we still have very few observational constraints on the progenitors of 
planets found around mature stars.  

AU~Mic, a member of the $\beta$~Pic moving group \citep[aged $\simeq$20~Myr,][]{Mamajek14,Miret20}, is a key object in this respect, being one of the closest and brightest pre-main sequence (PMS) 
M dwarfs.  Known to be hosting an extended debris disc with moving features \citep{Kalas04, Boccaletti15, Boccaletti18} and two transiting warm Neptunes \citep{Plavchan20, Martioli21},
AU~Mic is an ideal target for studying the formation and evolution of young planets and their atmospheres \citep{Hirano20}.
It has been extensively monitored with both photometry and spectroscopy \citep{Klein21,Cale21,Klein22,Zicher22,Szabo22,Donati23,Wittrock23,Mallorquin24,Yu25,Boldog25}, to further characterize 
the two transiting planets (called AU~Mic~b and c) and search for additional ones in the system.  Besides, AU~Mic is also known for its intense activity and strong magnetic field 
\citep{Kochukhov20,Klein21, Donati23}, making it a prime target for studying dynamos of mostly convective stars, magnetized winds and star-planet interactions \citep{Kavanagh21, 
Klein22, Alvarado22}, or escaping planetary atmospheres \citep{Hirano20,Carolan20,Allart23,Masson24}.

We had carried out a multi-season monitoring campaign of AU~Mic from early 2019 to mid 2022 \citep{Donati23} with the SPIRou nIR spectropolarimeter / high-precision velocimeter \citep{Donati20} 
at CFHT, mostly within the SPIRou Legacy Survey (SLS), a Large Programme of 310~nights with SPIRou focussing on planetary systems around nearby M dwarfs and on the study of planet formation around 
magnetically active stars. This spectropolarimetric and velocimetric monitoring allowed us to document the year-to-year evolution of the large- and small-scale magnetic field of AU~Mic.  It also 
enabled us to further characterize its planetary system, with the detection of a candidate outer planet at a distance of 0.17~au the existence of which was recently challenged by new observations 
in the optical domain, where activity is much stronger \citep{Mallorquin24}.  

In this paper we present follow-up monitoring of AU~Mic from mid 2022 to late 2024 with SPIRou, first within SPICE, a Large Programme of 174~nights at CFHT aiming at consolidating and
enhancing the results of the SLS until mid 2024, then with a PI programme in semester 2024B (PI: J.-F Donati, runIDs: 24BF15 and 24BF96), extending the time coverage to  
2041~d (5.6~yr).  We outline these new observations in Sec.~\ref{sec:obs}, describe the modeling of the time series with Gaussian Process Regression (GPR) in Sec.~\ref{sec:bls}, model all 
spectropolarimetric data of AU~Mic with Zeeman-Doppler imaging (ZDI) in a homogeneous way in Sec.~\ref{sec:zdi}, and carry out a new velocimetric analysis of the radial velocity 
(RV) data collected since the beginning of this monitoring effort in Sec.~\ref{sec:rvs}.  An overview of activity indexes is presented in Sec.~\ref{sec:act}.  We summarize the results in 
Sec.~\ref{sec:dis} and discuss their implications on the characterization of AU~Mic and its planetary system as well as the more general understanding of star and planet formation.

\section{New SPIRou observations}
\label{sec:obs}

During this monitoring program, we recorded another 161 spectra of AU~Mic with SPIRou, 18 in late 2022, 76 in 2023 and 67 in 2024.  For four of them (on 2023 May 03, and 2024 Nov 19, 23 and 25), 
poor and / or irregular weather yielded lower quality spectra with signal to noise ratios (SNRs) per 2.3~\kms\ pixel below 200, which we discarded from the analysis.  It yielded a 
total of 157 spectra (18, 75 and 64 in late 2022, 2023 and 2024, respectively) with SNRs ranging from 308 to 954 (median 825).  These 157 new spectra were added to the 225 older ones collected in  
the previous monitoring \citep{Donati23}, yielding a total of 382 spectra covering a period of 2041~d.  

We recall that SPIRou records spectra covering the entire 0.95--2.50~\mic\ (YJHK) wavelength range in a single exposure, at a resolving power of 70,000 \citep{Donati20}.  All spectra of AU~Mic were 
obtained in circular polarization (Stokes $V$) mode, with SPIRou polarization sequences consisting of four sub-exposures, each associated with a different orientation of the Fresnel rhomb retarders 
\cite[to remove systematics in polarization spectra to first order, see][]{Donati97b}.  Each recorded sequence yields one Stokes $I$ and one Stokes $V$ spectrum, as well as one null polarization 
check (called $N$) used to diagnose potential instrumental or data reduction issues.  In case of unstable weather where only 2 sub-exposures can be recorded, one can still retrieve
a pair of Stokes $I$ and $V$ spectra, but no $N$ spectrum and a slightly worse compensation of systematics.  Total exposure times per visit ranged from 379 to 802~s (median 802~s).  

All data were homogeneously processed with \texttt{Libre ESpRIT}, the nominal reduction pipeline of ESPaDOnS at CFHT adapted for SPIRou \citep{Donati20}.  These reduced spectra were used in 
particular for the spectropolarimetric analyses outlined in Secs.~\ref{sec:bls} and \ref{sec:zdi}.  Least-Squares Deconvolution \citep[LSD,][]{Donati97b} was applied to all reduced spectra, 
using a line mask constructed from the VALD-3 atomic and molecular line database \citep{Ryabchikova15} for an effective temperature \teff=3750~K and a logarithmic surface gravity \logg=4.5 adapted to AU~Mic, 
and selecting only atomic lines deeper than 10~percent of the continuum level, for a total of $\simeq$1500 lines of average wavelength and Land\'e factor equal to 1750~nm and 1.2 respectively.
The noise levels $\sigma_V$ in the resulting Stokes $V$ LSD profiles range from 0.74 to 2.16 (median 0.95)  in units of $10^{-4} I_c$ where $I_c$ is the continuum intensity.  

From these reduced spectra, we derived a homogeneous set of Stokes $I$ and $V$ LSD profiles and corresponding measurements of the longitudinal field \Bl, i.e., the line-of-sight-projected
component of the vector magnetic field averaged over the visible hemisphere, following \citet{Donati97b}.  The Stokes $V$ LSD signatures of AU~Mic being quite broad, the first moment is computed 
over a domain of $\pm45$~\kms\ about the line center, whereas the equivalent width of the Stokes $I$ LSD profiles is simply estimated through a Gaussian fit (and found to be $\simeq$2~\kms).  
The reduced chi-square \chisqr\ of the \Bl\ data (with respect to $\Bl=0$) is equal to 401, implying that the large-scale field of AU~Mic is unambiguously detected.  
We also computed LSD profiles for $N$, yielding $\chisqr=1.1$, consistent with no spurious signal down to the noise level and thereby indicating no issues in the observation and reduction 
procedures, nor in the derivation of \Bl\ and corresponding error bars.  The inferred \Bl\ values range from $-239$ to 257~G (median 39~G) over the 2041~d of this monitoring period, with error bars 
from 3.6 to 10.8~G (median 4.7~G).  The corresponding log of \texttt{Libre ESpRIT} reductions and associated quantities is given in Table~\ref{tab:log}, with rotation cycles and phases computed 
with a rotation period of 4.86~d and an arbitrary reference barycentric Julian date of BJD0~$=2459000$. 

AU~Mic observations were also processed with \texttt{APERO} (v0.7.292), the latest version of the SPIRou reduction pipeline \citep{Cook22}, which is much better optimised in 
terms of RV precision than \texttt{Libre ESpRIT}.  The reduced spectra were then analysed by the line-by-line (\texttt{LBL}) technique \citep[v0.65,][]{Artigau22}, using the median APERO spectrum 
of AU~Mic itself as the reference, to compute precise RVs for the 344 nightly-averaged observations, with 154 new RVs added to the 190 older ones from the previous monitoring (and 3 of the 157 new 
spectra discarded by \texttt{APERO}).  These RVs were corrected from the spectrograph drifts using the Fabry-Perot spectrum that SPIRou records simultaneously with the stellar spectrum \citep{Donati20}.  
The RV changes of AU~Mic (with respect to its average RV of $-4.51$~\kms) range from $-120$ to 95~\ms\ throughout this monitoring period, with nightly-averaged error bars spanning from 0.6 to 5.7~\ms\ 
(1.5 to 3.7~\ms\ excluding the transit night of 2019 June 16 and the two worst quality RV points, median 2.4~\ms).  From the variation in the depths of spectral lines, 
\texttt{LBL} also derives a precise estimate of the change in effective temperature $dT$ over the visible stellar hemisphere \citep{Artigau24}, as a result of starspots appearing and disappearing as the 
star rotates and the surface brightness distribution evolves with time.  We find that $dT$ ranges from $-20.2$ to 25.5~K, with error bars spreading from 0.4 to 3.0~K (median 1.3~K).  

As in \citet{Donati23}, the nightly-averaged spectra were also analysed with ZeeTurbo for estimating the small-scale magnetic field <$B$> at the surface of AU~Mic and its temporal evolution 
\citep[following][]{Cristofari23}, with all non-magnetic stellar parameters frozen to the values derived in the previous study from the median spectrum of AU~Mic \citep[see Table~1 of][]{Donati23}.  
We obtain that <$B$> ranges from 2.31 to 3.08~kG (median 2.74~kG) with error bars of 0.030 to 0.065~kG (median 0.038~kG).  The log of \texttt{APERO} reductions and associated measurements 
is given in Table~\ref{tab:log2}.

\section{The magnetic field and temperature variations of AU~Mic}
\label{sec:bls}

As in \citet[][]{Donati23}, we employed the framework of \citet{Haywood14} and \citet{Rajpaul15} to carry out a quasi-periodic (QP) GPR fit to the \Bl\ values, arranged in a vector denoted $\bf y$.  
The QP covariance function $c(t,t')$ we used for this purpose is as follows: 
\begin{eqnarray}
c(t,t') = \theta_1^2 \exp \left( -\frac{(t-t')^2}{2 \theta_3^2} -\frac{\sin^2 \left( \frac{\pi (t-t')}{\theta_2} \right)}{2 \theta_4^2} \right) 
\label{eq:covar}
\end{eqnarray}
where $\theta_1$ is the amplitude (in G) of the Gaussian Process (GP), $\theta_2$ its recurrence period (measuring \Prot),
$\theta_3$ the evolution timescale on which the \Bl\ curve changes shape (in d), and $\theta_4$ a smoothing parameter
describing the amount of allowed harmonic complexity.  
We add a fifth hyper parameter called $\theta_5$ that describes the excess of uncorrelated noise needed to obtain the QP GPR fit to the \Bl\ data with the highest 
likelihood $\mathcal{L}$, defined by:
\begin{eqnarray}
2 \log \mathcal{L} = -n \log(2\pi) - \log|C+\Sigma+S| - y^T (C+\Sigma+S)^{-1} y
\label{eq:llik}
\end{eqnarray}
where $C$ is the covariance matrix for all observing epochs, $\Sigma$ the diagonal variance matrix associated with $y$, $S=\theta_5^2 J$ the contribution of the additional white noise
where $J$ is the identity matrix, and $n$ the number of data points.
We then use a Monte-Carlo Markov Chain (MCMC) process to explore the hyper parameter domain, yielding posterior distributions and error bars for
each.  We use here the same MCMC and GPR modeling tools as in previous studies \citep[e.g.,][]{Donati23,Donati23b,Donati24,Donati24b}.
The MCMC process is a conventional single chain Metropolis-Hastings scheme, typically carried out over a few $10^5$ steps, including the first
few $10^4$ steps as burn-in.  {\emn Convergence is checked with an autocorrelation analysis, verifying that the burn-in and main phase are more than 
10$\times$ longer than the autocorrelation lengths of all parameters.}. 
The marginal logarithmic likelihood $\log \mathcal{L}_M$ of a given solution is computed following \citet{Chib01} as in \citet[][]{Haywood14}.  

\begin{figure*}[ht!]
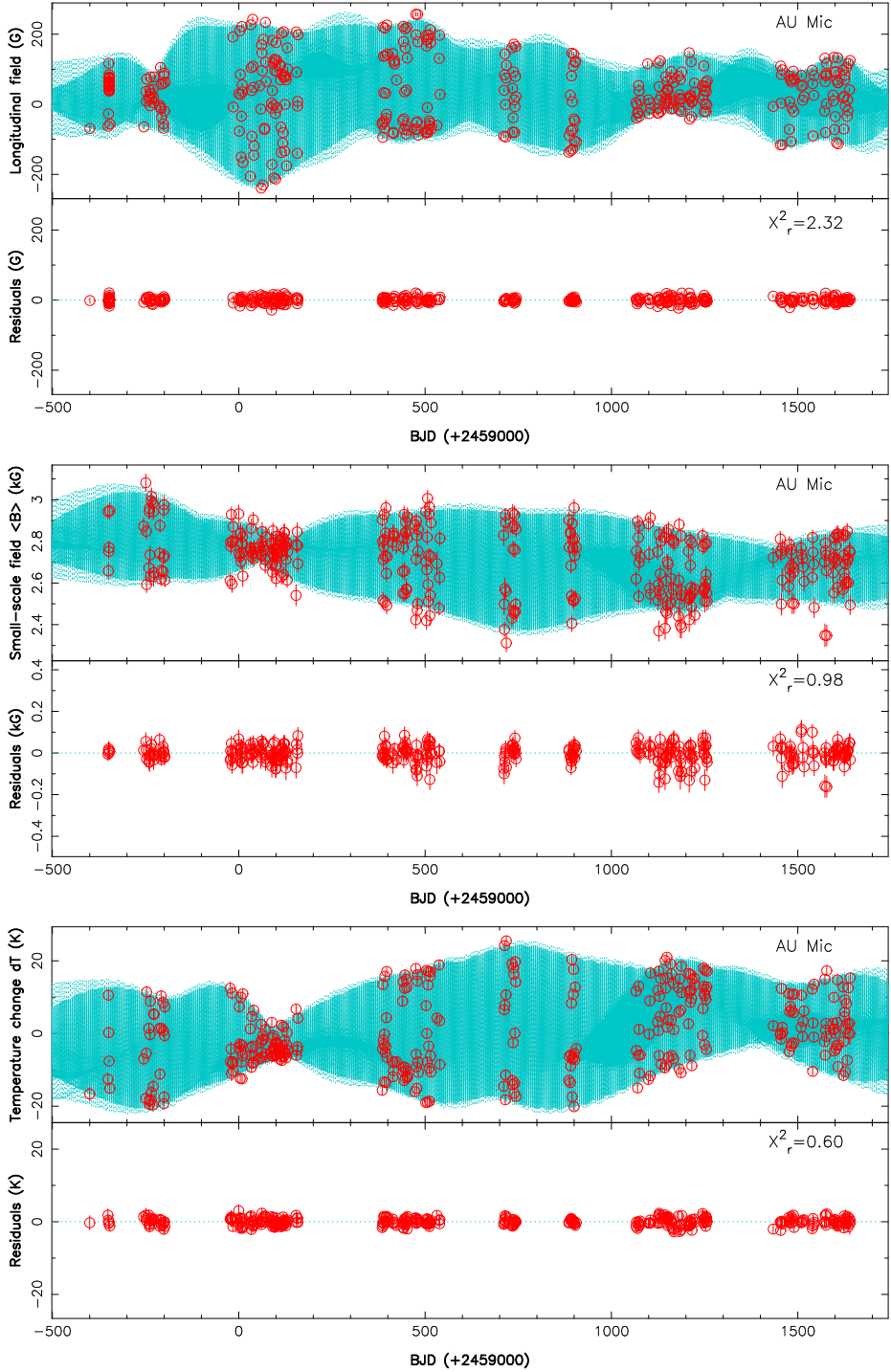

\centerline{\includegraphics[scale=0.47,angle=-90]{fignew/aumic2-gpb.ps}\vspace{2mm}}
\centerline{\includegraphics[scale=0.47,angle=-90]{fignew/aumic2-gpbf.ps}\vspace{2mm}} 
\centerline{\includegraphics[scale=0.47,angle=-90]{fignew/aumic2-gpt.ps}}
\caption[]{Longitudinal magnetic field \Bl\ (top panel), small-scale magnetic field <$B$> (medium panel) and temperature variations $dT$ (bottom panel) of AU~Mic (red dots), and QP GPR fit to the data 
(cyan full line) with corresponding 68~percent confidence intervals (cyan dotted lines).  The residuals, shown in the bottom plot of each panel, yield rms of 7.2~G,  0.038~kG and 0.96~K ($\chisqr=2.3$, 
0.98 and 0.60, respectively). {\emr A zoom on the 2023 and 2024 data is shown in Figs.~\ref{fig:gpb2} and \ref{fig:gpb3}.}}
\label{fig:gpb}
\end{figure*}

The result of the fit is shown in Fig.~\ref{fig:gpb}, with a zoom on the 2023 and 2024 data also provided in {\emr Figs.~\ref{fig:gpb2} and \ref{fig:gpb3}} 
\citep[see][for a zoom on the 2020 and 2021 data]{Donati23}.  The fitted GP parameters and error bars are listed in the top section of Table~\ref{tab:gpr}. 
All parameters are well defined, in particular the recurrence period $\theta_2$, found to be equal to $4.8591\pm0.0019$~d \citep[i.e., consistent within the error 
bar with the estimate of][]{Donati23} and the evolution timescale, which we measure at $101\pm9$~d, i.e., half the duration of a typical observing season.
The data are fitted to a rms level of 7.2~G, larger than the average error bar on the \Bl\ measurements (4.7~G), yielding $\chisqr=2.3$.
The \Bl\ data are thus not fitted down to the photon-noise level, suggesting an additional source of noise, such as intrinsic variability caused by
activity (e.g., stochastic changes in the large-scale field, flares), modeled by the GP models with $\theta_5$ being larger than 0.

The season-to-season variations of \Bl\ are detected very clearly, with the modulation amplitude shrinking to a minimum in October 2019 (BJD 2458800) and reaching a maximum in the following
season (BJD 2459100) before shrinking again to a minimum in 2023 (BJD 2460100).  Moreover, as expected from the rather short evolution timescale, the fitted \Bl\ curve also evolves 
significantly within each season (see Fig.~\ref{fig:gpb2}).  
Running GPR on individual seasons with dense coverage further suggests that evolution was faster in 2020 ($\theta_3=73\pm10$~d) than in 2021 ($\theta_3=140\pm45$~d) and 2023 
($\theta_3=160\pm45$~d), whereas the rotation period also fluctuates slightly from $4.8536\pm0.044$~d (in 2020) up to $4.8662\pm0.0045$~d (i.e., by about 3$\sigma$) as a likely 
result of differential rotation.  

As for \Bl, we carried out a QP GPR of the small-scale field <$B$>, and obtained the results shown in the middle panel of Fig.~\ref{fig:gpb} and 
the hyper-parameters listed in the middle section of Table~\ref{tab:gpr}, yielding $\chisqr=0.98$.  We also find that <$B$> is modulated by the 
rotation cycle, with a recurrence period of $4.8633\pm0.0016$~d, i.e., consistent at the 2$\sigma$ level with the period derived from \Bl.  The semi-amplitude of 
the <$B$> modulation is equal to $\theta_1=0.135\pm0.016$~kG on average, reaching a minimum of 0.1~kG in mid 2020 and a maximum of 0.3~kG in 2022.  We also note 
that <$B$> poorly (anti)correlates with \Bl\ (Pearson's coefficient $R=-0.4$) and even less with $|\Bl|$ ($R=-0.2$).  In fact, the modulation of <$B$> 
is found to be smallest when that of \Bl\ is largest (in 2020) and vice versa (in 2023).  This behaviour reflects that \Bl\ is very sensitive to 
the orientation of the local magnetic field vector whereas <$B$> is virtually not, yielding significantly different modulation patterns.  We 
also report that the evolution timescale is much longer for <$B$> than for \Bl, by a factor of about 2, partly reflecting that the temporal 
evolution of <$B$> is of much lower amplitude (with respect to the measurement error bar) and thus less precisely characterized than that of \Bl.  
Finally, we note that <$B$> decreases from an average of 2.81~kG at the beginning of the observations, down to about 2.64~kG in 2023, before 
starting to increase slightly (to 2.68~kG) in the last season.  


We also carried out the same analysis for the temperature change $dT$ of AU~Mic, yielding the results shown in the bottom panel of Fig.~\ref{fig:gpb} 
and the hyper-parameters in the bottom section of Table~\ref{tab:gpr}.  We find in particular that $dT$ is rotationally modulated,
with a period of $4.8611\pm0.0015$~d again consistent at the 2$\sigma$ level with the periods obtained from both \Bl\ and <$B$>.  The semi amplitude in the change 
of effective temperature (resulting from surface spots appearing and disappearing as the star rotates) is equal to $\theta_1=10.5\pm1.2$~K on average, 
varying from $\simeq$5~K in 2020 up to almost 25~K in 2022.  The evolution timescale is 1.6$\times$ longer than for \Bl\ and better defined than that for 
<$B$>, reaching $167\pm11$~d, demonstrating that topological changes in the large-scale field occur faster than changes in the location and intensity / 
contrast of the small-scale field and associated surface brightness features.  Besides, we obtain that, as already pointed out in \citet{Artigau24} on 
a smaller subset, $dT$ mimics <$B$> quite closely, the latter being strongly anti-correlated with the former at a level of $R=-0.92$.  This implies that $dT$ 
is a reliable proxy for <$B$> in the particular case of AU~Mic (and for a few other M dwarfs, Cristofari et al, submitted).  The measured change rate between 
<$B$> and $dT$ is $-76$~K/kG, with an intercept of $+$210~K for no magnetic field, suggesting that AU~Mic, if unspotted and non magnetic, would have a 
temperature about 210~K higher, i.e., $\teff\simeq3875$~K.  Assuming a typical spot-to-photosphere temperature contrast of 620~K for AU~Mic \citep{Berdyugina05}, 
we infer that the average fraction of the star covered with spots is $210/620=34$~percent.  The average $dT$ is minimum at the start of the observations 
($-5.5$~K) and rises until the penultimate season (reaching 5.5~K), then decreases in 2024 (to 2.5~K), implying seasonal changes in the spot coverage of 
AU~Mic similar to those induced by rotational modulation, i.e., $\pm$1$-$2~percent.

\section{Zeeman-Doppler imaging of AU~Mic}
\label{sec:zdi}

For this study, we carried out a complete re-analysis of the Stokes $IV$ data sets of AU~Mic over the full timescale of the monitoring.  
We started by splitting data from each season into 2 subsets more or less corresponding to A and B semesters (except for 2019 for which we only 
collected enough data for a dense phase coverage in semester B), yielding a total of 11 subsets (from 2019B to 2024B) spanning between 4 and 
25 rotation cycles of AU~Mic (i.e., 20 to 120~d, median 70~d).  Note that some of the 2023B data are also included in another magnetometric study of 
AU~Mic, exploiting additional Stokes $QU$ observations as well (Donati et al., in prep).  

We used the same ZDI code as in previous analyses \citep[e.g.,][]{Donati23}, which allows one
to reconstruct the relative photospheric brightness distribution and the topology of the large-scale magnetic field at the surface of a rotating star 
from phase-resolved sets of Stokes $I$ and $V$ LSD profiles.  This is achieved through an iterative process, starting from a small magnetic field and 
a featureless brightness map and progressively adding information to the surface of the star, exploring the parameter space until the modeled Stokes 
$I$ and $V$ profiles match the observed ones at the required level, usually $\chisqr\simeq1$ \citep[see, e.g.,][for more information on ZDI]{Brown91, 
Donati97c, Donati06b}.  In practice, the surface of the star is described as a grid of 5000 cells.  Whereas the relative photospheric brightness is 
simply described as a series of independent pixels, the large-scale magnetic field is expressed as a spherical harmonics expansion, using the formalism 
of \citet[][see also \citealt{Lehmann22,Finociety22,Donati23}]{Donati06b} in which the poloidal and toroidal components of the vector field depend on 3 
sets of complex coefficients, $\alpha_{\ell,m}$ and $\beta_{\ell,m}$ for the poloidal component, and $\gamma_{\ell,m}$ for the toroidal component, 
where $\ell$ and $m$ note the degree and order of the corresponding spherical harmonic term in the expansion.  As in Donati et al.\ (in prep), we used 
a spherical harmonic expansion with terms up to $\ell=10$, which is sufficient given the moderate rotational broadening of spectral lines 
\citep[$\vsini=8.5\pm0.2$~\kms,][]{Donati23} in the spectrum of AU~Mic.  As this inversion problem is ill-posed, ensuring a unique solution mandatorily 
requires regularization, which in the present case follows the principles of maximum entropy image reconstruction \citep{Skilling84} to select the image 
featuring minimal information among those matching the data.   In these ZDI reconstructions, we again assumed an inclination angle $i=80\degr$ between 
the rotation axis and the line of sight, i.e., slightly lower than the inclination of the orbital planes of planets b and c \citep[e.g.,][]{Szabo22}, 
to reduce mirroring effects of the imaging process between the upper and lower hemispheres.  

\begin{figure*}[ht!]
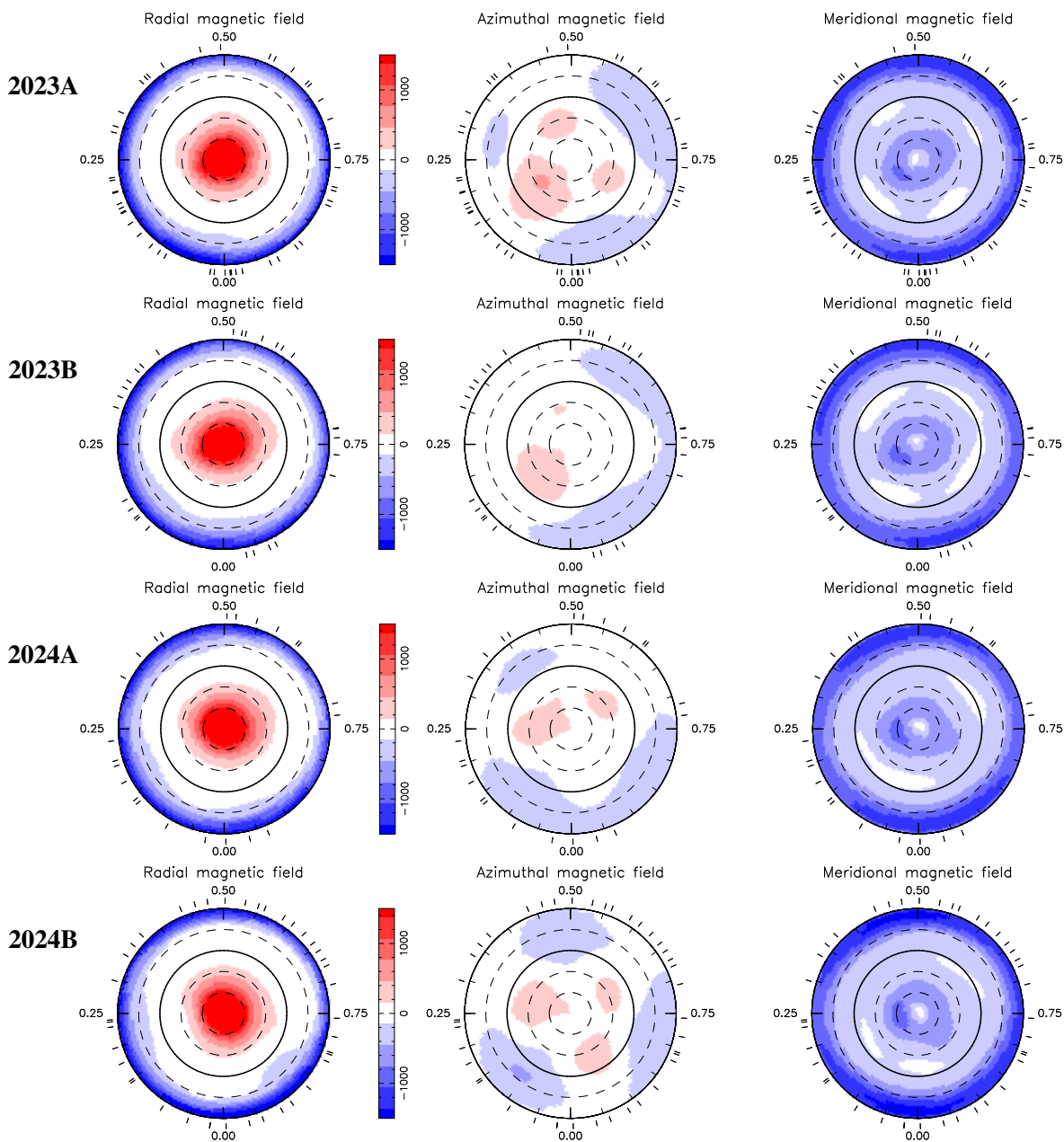

\centerline{\large\bf 2023A\raisebox{0.3\totalheight}{\includegraphics[scale=0.40,angle=-90]{fignew/aumic2-map23a.ps}}\vspace{1mm}}
\centerline{\large\bf 2023B\raisebox{0.3\totalheight}{\includegraphics[scale=0.40,angle=-90]{fignew/aumic2-map23b.ps}}\vspace{1mm}}
\centerline{\large\bf 2024A\raisebox{0.3\totalheight}{\includegraphics[scale=0.40,angle=-90]{fignew/aumic2-map24a.ps}}\vspace{1mm}}
\centerline{\large\bf 2024B\raisebox{0.3\totalheight}{\includegraphics[scale=0.40,angle=-90]{fignew/aumic2-map24b.ps}}} 
\caption[]{Reconstructed maps of the large-scale field of AU~Mic (left, middle and right columns for the radial, azimuthal and 
meridional components in spherical coordinates, in G), for season 2023A, 2023B, 2024A and 2024b (top to bottom row respectively), derived 
from the Stokes $V$ LSD profiles of Fig.~\ref{fig:fit} using ZDI.  The maps are shown in a flattened polar projection down to latitude
$-60$\degr, with the north pole at the center and the equator depicted as a bold line.  Outer ticks indicate phases of observations.
Positive radial, azimuthal and meridional fields respectively point outwards, counterclockwise and polewards. }
\label{fig:map}
\end{figure*}

To compute local synthetic Stokes $IV$ profiles from each grid cell, we use Unno-Rachkovsky's analytical solution of the polarized radiative transfer 
equation in a plane-parallel Milne Eddington atmosphere \citep{Landi04}.  We then integrate the spectral contributions from all visible grid cells 
(assuming a linear center-to-limb darkening law for the continuum, with a coefficient of 0.3) to obtain the global synthetic profiles at each observed 
rotation phase.  The mean characteristics of the synthetic profiles are copied from those of the observed LSD profiles, i.e., a central wavelength of 
1700~nm, a Land\'e factor of 1.2 and a Doppler width of $\vD=3.5$~\kms\ \citep[as in][]{Donati23}.  We introduce a first filling factor $f_V$ 
(assumed constant over the whole star) describing the fraction of each grid cell that contributes to the large-scale field and to Stokes $V$ profiles, 
implying a magnetic field of $B_V/f_V$ in the magnetic portion of the cell and a magnetic flux of $B_V$ over the whole cell.  Similarly, we 
assume that a fraction $f_I$ of each grid cell (called the filling factor of the small-scale field, again equal for all cells) hosts small-scale fields of 
strength $B_V/f_V$, implying a small-scale magnetic flux over the whole cell of $B_I = B_V f_I/f_V$.  In this context, <$B$> measured with ZeeTurbo 
at a given epoch is equal to the weighted limb-darkened average of $B_I$ over the visible stellar hemisphere.  This simple parametric approach within ZDI, 
used to empirically reproduce the co-existence of small-scale and large-scale fields, was found to provide an adequate 
description of the Stokes $IV$ data sets of AU~Mic \citep[][Donati et al.\ in prep.]{Donati23}.  In this paper, we chose not to apply ZDI on 
Stokes $V$ profiles only, given the strong underestimation of both the large- and small-scale field that this approach leads to \citep[][]{Donati23}.  
We also chose not to attempt estimating surface differential rotation, occurring on a timescale comparable 
to the evolution of the large-scale field and thus hard to reliably detect with ZDI \citep{Donati23}.  

\begin{figure}[ht!]
\centerline{\includegraphics[scale=0.3,angle=-90]{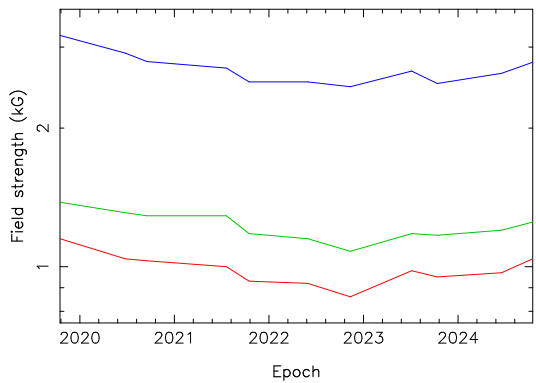}}
\caption[]{Quadratic average of the large-scale magnetic field over the stellar surface (red, column 2 of Table~\ref{tab:mag}), polar field strength of the 
dipolar component (green, column 5 of Table~\ref{tab:mag}) and average small-scale field over the rotation cycle (blue, column 4 of Table~\ref{tab:mag}) as 
a function of the observing epoch, for the 11 magnetic topologies of AU~Mic reconstructed with ZDI. } 
\label{fig:zdib}
\end{figure}

We applied ZDI to the 11 data subsets, and we show the achieved fits to the LSD Stokes $I$ and $V$ profiles in Fig.~\ref{fig:fit}, and the recovered magnetic 
maps in Fig.~\ref{fig:map} for the last 4 subsets (2023A to 2024B).  The reconstructed maps in previous semesters are showed 
in Figs.~\ref{fig:map2} and \ref{fig:map3}.  Photospheric brightness maps were also simultaneously reconstructed as part of the 
imaging process, but show only few low-contrast features and are thus not displayed in Fig.~\ref{fig:map}.  On stars with low \vsini, ZDI indeed reconstructs 
mostly the non-axisymmetric component of the brightness distribution, inducing the observed modulation of Stokes $I$ LSD profiles and covering only a few 
percent of the stellar surface (in agreement with the observed modulation of $dT$, see Sec.~\ref{sec:bls}).  
As described in the previous study, fitting the LSD Stokes $I$ and $V$ profiles at the same time allows one to obtain a more reliable description of both the 
large-scale and small-scale fields of AU~Mic, with respect to a reconstruction exploiting Stokes $V$ data only that suffers more cancellation from the 
northern and southern hemisphere given the almost equator-on orientation of AU~Mic.  The epoch-to-epoch variations are real, as was clear from the \Bl\ 
and <$B$> times series (see Sec.~\ref{sec:bls} and Fig.~\ref{fig:gpb}, although not so obvious to diagnose visually from the maps 
themselves.  We list in Table~\ref{tab:mag} the quantitative characteristics of the reconstructed magnetic topologies for each of the 11 subsets, and 
display the main ones as a function of the observing epoch in Fig.~\ref{fig:zdib}.  
We find in particular that the large-scale field progressively weakens from 2019B to 2022B before increasing till the end of the campaign.  
The inferred magnetic topology, almost fully poloidal and axisymmetric at all epochs, features a dominant 1.1 to 1.4~kG dipole component inclined at 
10$-$20\degr\ to the rotation axis.  This dipolar component contains $\simeq$70~percent of the reconstructed magnetic energy all all epochs, and varies both in strength (minimum in 2022B) 
and orientation with time.  The average small-scale field <$B_s$> that we derive from these reconstructions (from a weighted average over the visible 
stellar hemisphere, taking into account limb-darkening, see column 4 of Table~\ref{tab:mag}) predicts slightly too large an epoch-to-epoch variation.   
Otherwise it is consistent with observations (decreasing from 2019 to 2022-2023 before increasing in 2024, see Fig.~\ref{fig:gpb}), and with 
the full-amplitude modulation at each epoch (minimum in 2020 and maximum in 2022).  Besides, ZDI suggests that the largest Zeeman broadening 
comes from the strongest field near the poles, similar to previous results from Stokes $I$ data on another young active star \citep{Kochukhov23}.  

We note that attempting to simultaneously fit all data collected in a single season (e.g., 2024A and 2024B) always yields a larger \chisqr\ and thus a 
significantly worse fit than that obtained when splitting each season in half, temporal evolution (including differential rotation) 
becoming quite notable on a timescale of up to 207~d (in 2024).  In fact, Stokes $V$ LSD profiles within the half-season subsets already show evidence 
for temporal variability beyond rotational modulation, with slightly discrepant Zeeman signatures detected at almost equal rotational phases but 
different cycles (e.g., 12.336 and 20.354 in 2024, see bottom right panel of Fig.~\ref{fig:fit}).  This is actually expected given the evolution 
timescale $\theta_3$ derived from the \Bl\ analysis of Sec.~\ref{sec:bls} ($101\pm9$~d, see Table~\ref{tab:gpr}), comparable to the maximum length of 
the subsets but much shorter than the duration of the observing seasons.  
We also note that the field strengths reconstructed in this new study are about twice as large as those presented in the previous study \citep{Donati23} 
at the same epochs, mostly reflecting the much tighter fit to the Stokes $I$ profiles that we were able to achieve in this new work.  As a result, the 
new inferred magnetic topologies yield small-scale field values (through a weighted average of the reconstructed field strengths over the visible 
hemisphere, see above) that are much more consistent with actual measurements, and agree well with the results of the latest magnetometric study of 
AU~Mic from Stokes $IVQU$ data collected in 2023B (Donati et al., in prep).

\section{Radial velocities of AU~Mic}
\label{sec:rvs}

In this section we simultaneously analyse the full set of RVs derived for AU~Mic from 2019 to 2024, repeating most of the previous analysis on 
the full set of 344 measurements now spanning 2041~d in the life of the AU~Mic multi-planet system.  As mentioned in Sec.~\ref{sec:obs}, the new RVs are derived 
with the latest versions of \texttt{APERO} and \texttt{LBL}, implementing in particular a more accurate correction of telluric contamination and expected to yield 
even more precise RV estimates. 

Following the previous study \citep{Donati23}, we analysed three main cases, a reference one including transiting planets b and c, a second one including b, c and the candidate 
outer planet e \citep[the existence of which was recently challenged,][]{Mallorquin24}, and a third one including b, c, e and the putative planet d located between b and c 
and presumably causing the reported transit time variations (TTVs) of planets b and c \citep{Wittrock23,Boldog25}.  For each of these three cases, we fit the observed RV curve 
with a model including a dominant activity signal from the star, described with a QP GP (as in Sec.~\ref{sec:bls}), as well as a Keplerian signal for the considered planets.  
Although we know that transiting planets b and c truly exist, we nonetheless included a fourth case where no planets are included to determine how well their RV signatures 
are detected and to compare with the previous results.  In each case, the free model parameters and corresponding error bars are inferred 
through a MCMC process (as in Sec.~\ref{sec:bls}).  

\begin{table*}[ht!]
\caption{MCMC results for the four studied cases (no planet, b+c, b+c+e, b+c+e+d)} 
\centering
\resizebox{0.8\textwidth}{!}{
\begin{tabular}{cccccc}
\hline
Parameter          & No planet                 & b+c                       & b+c+e                     & b+c+e+d                   &   Prior \\ 
\hline
$\theta_1$ (\ms)   & $37.6^{+4.1}_{-3.7}$ & $38.0^{+4.1}_{-3.7}$ & $39.1^{+4.4}_{-4.0}$ & $39.0^{+4.5}_{-4.0}$ & mod Jeffreys ($\sigma_{\rm RV}$) \\  
$\theta_2$ (d)     & $4.8654\pm0.0015$    & $4.8650\pm0.0015$    & $4.8652\pm0.0014$    & $4.8652\pm0.0014$    & Gaussian (4.86, 0.1) \\ 
$\theta_3$ (d)     & $167^{+15}_{-14}$    & $166^{+14}_{-13}$    & $169^{+15}_{-14}$    & $168^{+15}_{-14}$    & log Gaussian ($\log$ 160, $\log$ 1.5) \\ 
$\theta_4$         & $0.33\pm0.03$        & $0.32\pm0.03$        & $0.35\pm0.03$        & $0.35\pm0.03$        & Uniform  (0, 3) \\ 
$\theta_5$ (\ms)   & $12.0\pm0.6$         & $12.3\pm0.6$         & $11.9\pm0.6$         & $12.0\pm0.6$         & mod Jeffreys ($\sigma_{\rm RV}$) \\ 
\hline
$K_b$ (\ms)        &                      & $2.8^{+1.1}_{-0.8}$  & $2.7^{+1.1}_{-0.8}$  & $2.8^{+1.1}_{-0.8}$  & mod Jeffreys ($\sigma_{\rm RV}$) \\ 
$P_b$ (d)          &                      & 8.463446             & 8.463446             & 8.463446                  & fixed from \citet{Mallorquin24} \\ 
BJD$_b$ (2459000+) &                      & $-669.649175$        & $-669.649175$             & $-669.649175$             & fixed from \citet{Mallorquin24} \\ 
$M_b$ (\me)        &                      & $6.3^{+2.5}_{-1.8}$  & $6.1^{+2.5}_{-1.8}$ & $6.3^{+2.5}_{-1.8}$ & derived from $K_b$, $P_b$ and \mstar \\ 
\hline
$K_c$ (\ms)        &                      & $4.1^{+1.1}_{-0.9}$  & $3.9^{+1.1}_{-0.9}$  & $3.8^{+1.1}_{-0.9}$  & mod Jeffreys ($\sigma_{\rm RV}$) \\ 
$P_c$ (d)          &                      & 18.859018            & 18.859018            & 18.859018                 & fixed from \citet{Mallorquin24} \\ 
BJD$_c$ (2459000+) &                      & $-657.776513$        & $-657.776513$        & $-657.776513$             & fixed from \citet{Mallorquin24} \\ 
$M_c$ (\me)        &                      & $12.1^{+3.3}_{-2.7}$ & $11.6^{+3.3}_{-2.7}$ & $11.3^{+3.3}_{-2.7}$ & derived from $K_c$, $P_c$ and \mstar \\ 
\hline
$K_e$ (\ms)        &                      &                           & $5.9^{+1.5}_{-1.2}$  & $5.8^{+1.5}_{-1.2}$  & mod Jeffreys ($\sigma_{\rm RV}$) \\  
$P_e$ (d)          &                      &                           & $33.11\pm0.06$       & $33.10\pm0.06$       & Gaussian (33.1, 1.0) \\ 
BJD$_e$ (2459000+) &                      &                           & $118.5\pm1.4$        & $118.6\pm1.4$        & Gaussian (118, 8) \\ 
$M_e$ (\me)        &                      &                           & $21.1^{+5.4}_{-4.3}$ & $20.7^{+5.4}_{-4.3}$ & derived from $K_e$, $P_e$ and \mstar  \\ 
\hline
$K_d$ (\ms)        &                 &                           &                           & $1.0^{+0.9}_{-0.5}$  & mod Jeffreys ($\sigma_{\rm RV}$) \\  
$P_d$ (d)          &                 &                           &                           & 12.73596             & fixed from \citet{Wittrock23} \\ 
BJD$_d$ (2459000+) &                 &                           &                           & $-659.44219$         & fixed from \citet{Wittrock23} \\ 
$M_d$ (\me)        &                 &                           &                           & $2.6^{+2.3}_{-1.3}$ & derived from $K_d$, $P_d$ and \mstar  \\ 
\hline
\chisqr            & 21.9                 & 19.6                 & 18.4                 & 18.3                 &  \\ 
rms (\ms)          & 11.0                 & 10.4                 & 10.1                 & 10.1                 &  \\ 
$\log \mathcal{L}_M$ & 878.1              & 890.5                & 902.0                & 902.3                &  \\
$\log {\rm BF} = \Delta \log \mathcal{L}_M$ & $-12.4$  & 0.0     & 11.5                 & 11.8                 &  \\
\hline 
\end{tabular}}
\tablefoot{\emr In each case, we list the recovered GP and planet parameters with their error
bars, as well as the priors used whenever relevant.  The last 4 rows give the \chisqr\ and the rms of the best fit to the RV data, as well as the associated marginal
logarithmic likelihood $\log \mathcal{L}_M$ and marginal logarithmic likelihood variation $\Delta \log \mathcal{L}_M$ with respect to the reference case (b+c).}
\label{tab:pla}
\end{table*}

\begin{figure*}[ht!]
\centerline{\includegraphics[scale=0.58,angle=-90]{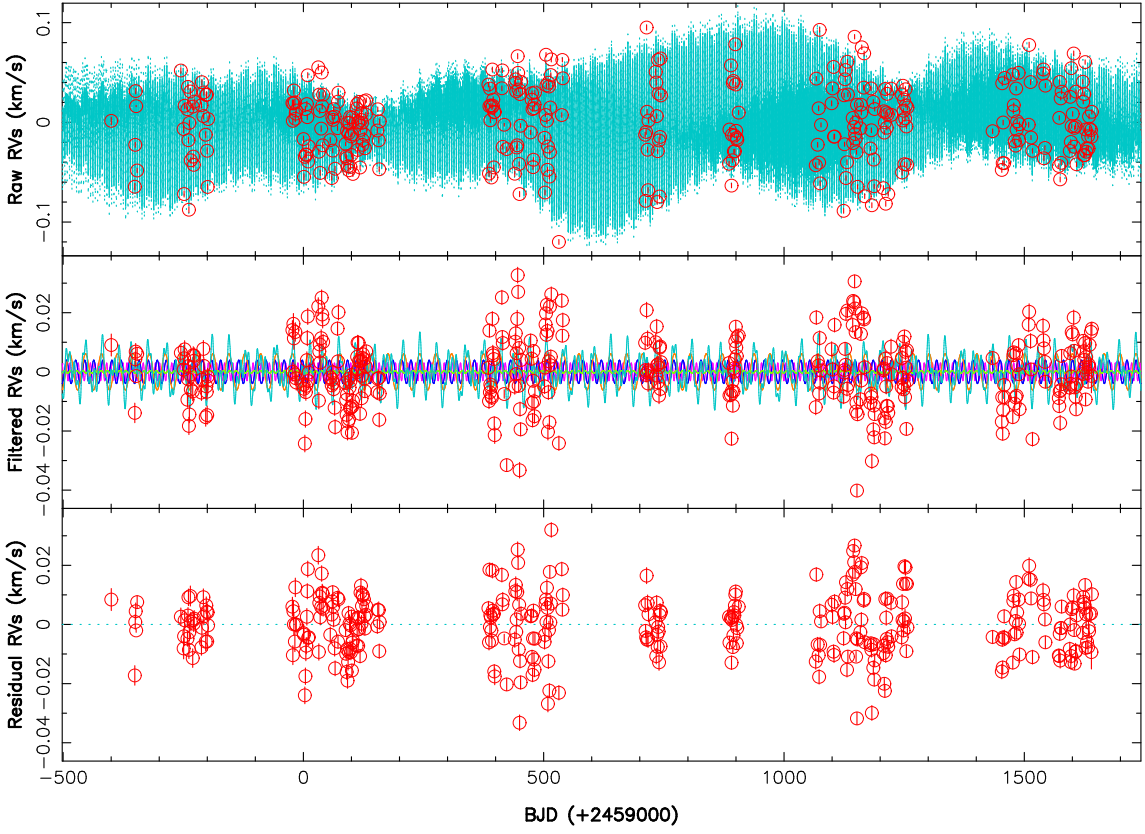}}
\caption[]{Raw (top), filtered (middle) and residual (bottom) RVs of AU~Mic (red dots) over the observing period.  The top plot shows the MCMC fit to the data, including a QP GPR modeling of the activity and
the RV signatures of all four planets (cyan), whereas the middle plot shows the planet RV signatures (pink, blue, green, orange and cyan for planets b, c, d, e and b+c+d+e, respectively) once activity 
is filtered out.  The rms of the residuals is 10.1~\ms.  {\emr A zoom on the 2023 and 2024 data is shown in Fig.~\ref{fig:rvr2}.}}
\label{fig:rvr}
\end{figure*}

Orbital periods and transit (or conjunction) times being already determined accurately for b, c and d from photometric and TTV data \citep[e.g.,][]{Szabo22,Wittrock23,Mallorquin24}, 
we chose to simply fix them for the analysis, which is virtually equivalent to let them vary within the very narrow prior determined from previous studies.  
Assuming that orbits of all planets are circular (as a first step), we are left with six planet parameters, the RV semi-amplitudes of all four planets (denoted $K_b$, $K_c$ $K_d$ and 
$K_e$), and the orbital period and conjunction time of planet e ($P_e$ and BJD$_e$).  Of these six planet parameters, we respectively fit zero, two ($K_b$, $K_c$), five (all but $K_d$) 
or all six from the RV data in the four considered cases, in addition to the five GP hyper parameters describing the activity.  
The values and error bars we derive with the MCMC process for the relevant parameters are listed in Table~\ref{tab:pla} (with the corresponding priors), whereas the best fit we achieved 
to the observed RVs is shown in Fig.~\ref{fig:rvr} {\emr (with a zoom on the 2023 and 2024 data in Fig.~\ref{fig:rvr2})} when all four planets are 
included.  We find that the model featuring no planet definitely yields a worse fit to the SPIRou RVs of AU~Mic than that including planets b and c ($\Delta \log \mathcal{L}_M=-12.4$).  
In the other cases, the average semi-amplitudes derived for planets b and c are equal to $K_b=2.8^{+1.1}_{-0.8}$~\ms\ and $K_c=3.9^{+1.1}_{-0.9}$~\ms, and correspond to masses of $M_b=6.3^{+2.5}_{-1.8}$~\me\ and 
$M_c=11.6^{+3.3}_{-2.7}$~\me\ (where \me\ is the Earth's mass), and to detection levels of 3.5 and 4.3$\sigma$.  These semi-amplitudes are smaller than, but still consistent within error bars 
with, those inferred from the previous study, and to the recent estimates from the joint analysis of CARMENES, HARPS and the first set of SPIRou data \citep[$K_b=3.6\pm1.1$~\ms\ and 
$K_c=4.3\pm1.0$~\ms,][]{Mallorquin24}.  

\begin{figure}[ht!]
\centerline{\includegraphics[scale=0.36,angle=-90]{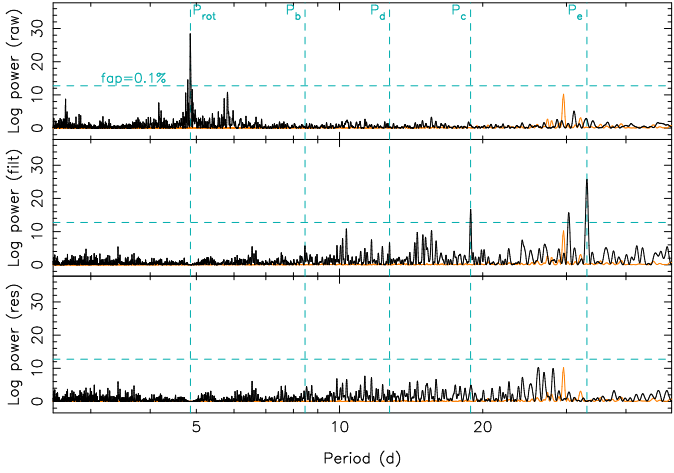}}  
\caption[]{Periodogram of the raw (top), filtered (middle) and residual (bottom) RVs when including all planets in the MCMC modeling.  
The cyan vertical dashed lines trace the rotation period of the star and the planet orbital periods, whereas the horizontal dashed line show the 0.1\% FAP level in the 
periodogram of the RV data.  The peak corresponding to candidate planet e (with a 1-yr alias at 30.3~d) dominates the middle plot, with a FAP of $2\times10^{-9}$.   
The orange curve depicts the window function, with a peak at the synodic period of the Moon (at 29.5~d).  } 
\label{fig:per} 
\end{figure}

\begin{figure}[ht!]
\centerline{\includegraphics[scale=0.32,angle=-90]{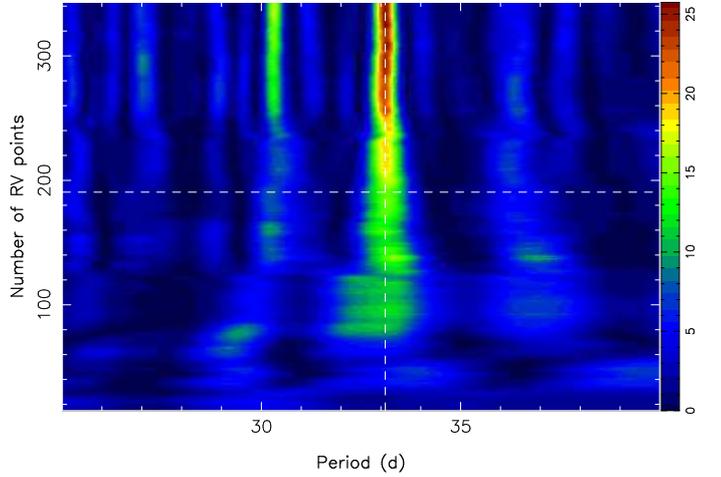}}
\caption[]{Stacked periodograms of the filtered RVs, as a function of the number of RV points included in the Fourier analysis (starting from the first collected spectrum), 
with a color scale depicting the logarithmic power in the periodogram.
The main RV signal associated with candidate planet e, outlined with a vertical dashed line (see Table~\ref{tab:pla}), gets stronger and increasingly dominant in this period range as more spectra are added to the analysis.  
The horizontal dashed line illustrates the end of the previous data set \citep{Donati23}.  The weaker peak at 30.3~d, also visible in Fig.~\ref{fig:per} (middle plot), is a 1-yr alias of the main one. }
\label{fig:stp}
\end{figure}

\begin{figure}[ht!]
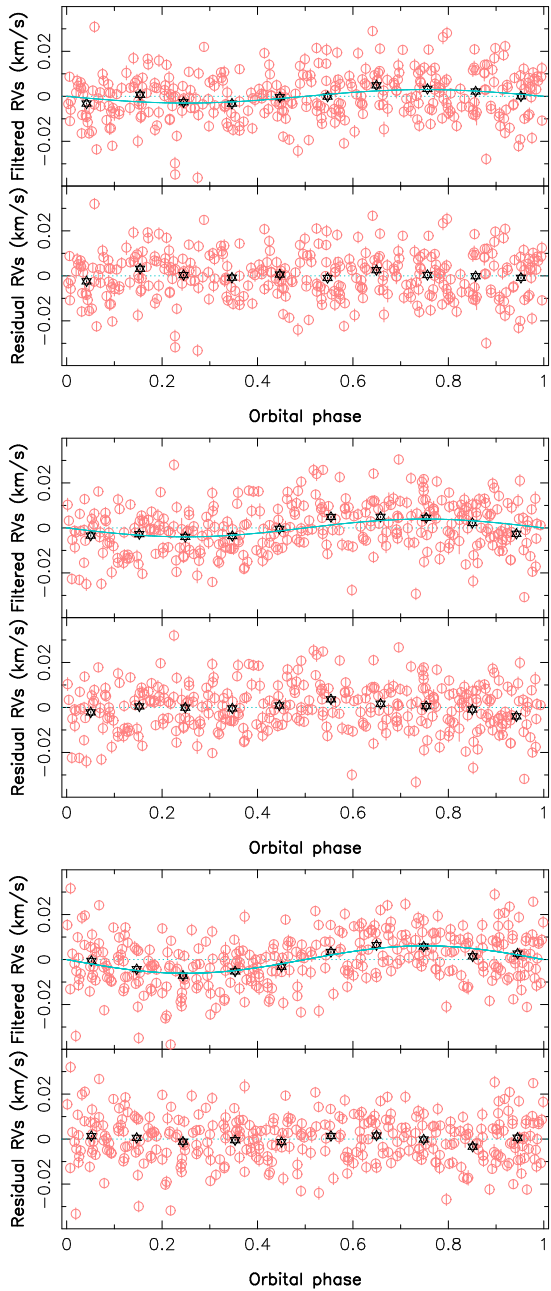

\centerline{\includegraphics[scale=0.45,angle=-90]{fignew/aumic2-rvb.ps}\vspace{1mm}}
\centerline{\includegraphics[scale=0.45,angle=-90]{fignew/aumic2-rvc.ps}\vspace{1mm}}
\centerline{\includegraphics[scale=0.45,angle=-90]{fignew/aumic2-rve.ps}}
\caption[]{Phase-folded filtered (top plots) and residual (bottom plots) RVs for transiting planets b (top panel) and c (middle panel), and for candidate planet e (bottom panel) of 
AU~Mic.  The red dots are the individual RV points with the respective error bars, whereas the black stars are average RVs over 0.1 phase bins.  
As in Fig.~\ref{fig:rvr}, the dispersion of RV residuals is 10.1~\ms.   } 
\label{fig:rvf}
\end{figure}

We also find that including candidate planet e provides a significant improvement in the fit ($\Delta \log \mathcal{L}_M=11.5$) although the derived semi-amplitude, equal to $K_e=5.9^{+1.5}_{-1.2}$~\ms, 
is about half as large as (though still compatible within about 2.5$\sigma$ with) that derived in the initial study \citep[][]{Donati23}.  It would imply that candidate planet e, if real, has a
mass of $M_e=21.1^{+5.4}_{-4.3}$~\me, and is detected at a level of 4.9$\sigma$.  At $P_e=33.11\pm0.06$~d, its orbital period is also slightly shorter (by about 0.3~d or 2.5$\sigma$) than the  
previous measurement, but still consistent with orbits that would potentially be stable on a Gyr timescale \citep[for small eccentricities, see Fig.~12 of][]{Donati23}, i.e., much longer than the 
age of the AU~Mic system, assuming all planets are coplanar.  
We show in Fig.~\ref{fig:per} the periodogram of the raw, activity-filtered and residual RVs, with the peak corresponding to candidate planet e associated with a false 
alarm probability (FAP) that the signal is spurious of about $2\times10^{-9}$ in the filtered RVs.  Fig.~\ref{fig:stp} presents a stacked periodogram of the filtered RVs showing a signal at $P_e$ that 
consistently strengthens (rather than varying up and down like a spurious signal from activity) as data are added to the analysis, suggesting that candidate planet e is real.  The updated value of 
$K_e$ is also significantly lower than the 3.5$\sigma$ upper limit of 10~\ms\ derived by \citet{Mallorquin24}, potentially explaining why they did not detect it.  The phase-folded RV curves of b, c 
and e are shown in Fig.~\ref{fig:rvf}.  Including candidate planet d only marginally improves the likelihood ($\Delta \log \mathcal{L}_M=0.3$) with respect to the b+c+e case, without significantly 
changing the results for the other three planets.  The 90 and 99~percent confidence upper limits we derive on $K_d$ are equal to 1.9 and 2.7~\ms\ (respectively 4.9 and 7.0~\me\ for $M_d$).  Running the 
same experiment with Keplerian (i.e., non circular) orbits for transiting planets b and c and for candidate planet e only yields a marginal improvement $\Delta \log \mathcal{L}_M\simeq1.5$, confirming 
a posteriori that circular orbits are the most likely scenario for all three planets, with respective error bars on the eccentricities equal to 0.02, 0.12 and 0.15 for b, c and e.  

Finally, we note that the hyper-parameters of the GP describing the activity of AU~Mic over the 2041~d of the observations are similar in all four cases, with an average semi-amplitude 
$\theta_1$ of about 40~\ms, i.e., 25 percent larger than in the previous study though still a factor of 2.5 to 4 smaller than that at optical and visible wavelengths \citep{Mallorquin24}.  
At about 167~d, the evolution timescale $\theta_3$ is also longer by about 20 percent than the older estimate, and fully consistent with that derived from $dT$ (see bottom section of 
Table~\ref{tab:gpr});  it is also much longer than the orbital periods of b, c, d and e, suggesting minimal interference between the GPR fit and the planet parameters.  The recurrence period 
$\theta_2$ (another proxy for \Prot) is marginally longer than that found from \Bl\ and $dT$ but consistent with that derived from <$B$>, all 
four inferred with very small error bars ranging from 0.0014 to 0.0019~d (2.0 to 2.7~min).  We also report slight seasonal changes in the GP parameters similar to those found for \Bl, 
in particular marginally shorter evolution timescale ($\theta_3=129\pm40$~d) and recurrence period ($\theta_2=4.850\pm0.007$~d) in 2020 than in the other seasons.

\section{Activity of AU~Mic}
\label{sec:act}

We have also investigated the temporal behaviour of a few additional activity proxies (beyond those already mentioned in Sec.~\ref{sec:bls}) that we briefly discuss below.  

In addition to providing precise RVs \citep[through the coefficient of the first term in the Taylor expansion used to describe the variational profile of each line with respect to its median, 
see][]{Artigau22}, \texttt{LBL} also yields estimates of the changes in line width and asymmetry (through the coefficients of the second and third terms in the same Taylor expansion,  
called $d2v$ and $d3v$), which share similarities with the more conventional proxies called differential line width and bisector span (dLW and BIS) computed from the cross-correlation profile 
of spectral lines.  Fitting $d2v$ and $d3v$ with GPR (as \Bl\ and $dT$ in Sec.~\ref{sec:bls}), we find that both quantities are rotationally modulated, with a recurrence period consistent 
within error bars with that derived from RVs.  We also obtain that $dT$ correlates reasonably well with $d2v$ ($R=0.66$) though less tightly than with <$B$> (see 
Sec.~\ref{sec:bls}), and that RVs are anti-correlated with $d3v$ ($R=-0.60$, as usually the case with BIS when RVs are dominated by the activity jitter).  
The best correlation with RVs we find is however neither with $d2v$ nor with $d3v$ (or their first time derivatives) but with the first time derivative of $dT$ ($R=-0.85$).  Even then, 
exploiting this correlation to filter RVs from the activity jitter still leaves us with a residual activity signal of semi-amplitude equal to about 40~percent that of the original activity 
signal \citep[similar to the results of the first study, see][]{Donati23}.  

We also tried a multi-dimensional GPR fit to our RV and $dT$ data simultaneously \citep[as advocated by][]{Rajpaul15} but still end up with a residual RV activity signal of semi-amplitude 
$\simeq$1/3 of the original one, {\emn likely reflecting the limitation of the $FF'$ approach underlying the multi-dimensional GPR modeling (see Tables~\ref{tab:pla} and \ref{tab:gpr}).  
We note that the multi-dimensional GP significantly reduced the GP evolution timescale (from $\simeq$170 to $\simeq$120~d) in an attempt to compensate for this limitation, but with moderate 
success.  The achieved fit to the RVs is significantly worse than with a simple GP, with a 2.4$\times$ larger \chisqr, a 1.55$\times$ higher rms and a 1.4$\times$ higher excess RV jitter.  
Despite this limitation, the parameters inferred for planets b, c and candidate planet e, are consistent with those derived in Sec.~\ref{sec:rvs} within $<$$1\sigma$ (albeit with larger error bars), 
and so are the corresponding $\log$~BF values with respect to the two-planet reference case (see results in Table~\ref{tab:pla2})}.  We thus conclude that, for stars with dominant RV activity signal 
like AU~Mic, straightforward GPR analyses of RVs remain the most efficient approach to filter densely sampled RV curves from the activity jitter.  

We also looked at the 1083~nm \hei\ and 1282~nm \pab\ lines, known as reliable chromospheric activity proxies in low-mass stars.  To investigate this point, we proceeded as in \citet{Donati23}, i.e., 
normalise each \hei\ and \pab\ profile by the corresponding median of all recorded profiles, then fit the normalized profile by a Gaussian of fixed full-width at half maximum (FWHM=40~\kms) centred 
on the stellar rest frame to find out how much each line changed with respect to the median at all epochs.  Despite AU~Mic being quite active, we find that both lines are only weakly variable, 
with equivalent width fluctuations of order 0.8~\kms\ rms for \hei\ and 0.3~\kms\ rms for \pab\ over the full set of observations.  Most of this dispersion is attributable to intrinsic variability, 
with a few major eruptive events occurring from time to time, e.g., on 2019 June 14, 2020 May 30, 2021 November 13, 2023 August 28 and 2024 September 22, during which both lines can show up in 
emission, with broad wings extending beyond $\pm100$~\kms\ in the stellar rest frame and equivalent widths exceeding 15~\kms.  Outside of flares, rotational modulation is weak, consistent with 
the fact that the magnetic topology is mostly axisymmetric, with average semi-amplitudes on the equivalent width variations of about 0.2 and 0.1~\kms\ for \hei\ and \pab\ respectively.  
Modulation is nonetheless strong enough to still dominate the periodogram of both lines over the full range of the observations.  We also checked that flares, even major ones, do not have 
a clear impact on the measured RVs (at a rms level of $\simeq$10~\ms), with residuals to the GPR fit that are consistent with the bulk of the data points, and a GPR fit excluding RVs at 
flaring epochs yielding results consistent within 0.5$\sigma$ with that including all RV points.  

Focussing on the fluctuations of the \hei\ line, more clearly detected than those of \pab, and computing a 2D periodogram for the 11 subsets outlined in Sec.~\ref{sec:zdi}, we find that there is 
a dominant peak at \Prot\ in subset 2019B only (see Fig.~\ref{fig:eml}, left panel) but in no other subsets.  Otherwise \Prot\ is at best weak (as in, e.g., 2023B) and most of the time undetected, even 
when spectra featuring large flares or those most strongly contaminated with telluric lines are left out, reflecting the high level of intrinsic variability with respect to rotational modulation.  
In 2019B, maximum \hei\ emission occurs at phase 0.92, i.e., slightly before the positive magnetic 
pole gets furthest from the observer (at phase 0.02, see Table~\ref{tab:mag}), as expected from the magnetic equator being brighter than the pole at the chromospheric level.  We also note one subset 
(2020B) where significant power is detected at the orbital period of planet b (see Fig.~\ref{fig:eml}, right panel), potentially suggesting energetic star-planet interactions at this particular 
epoch with maximum \hei\ emission occurring 2.4~d after the transit.  
We however stress that this detection, although showing up across the whole \hei\ triplet as expected from a real signal, may actually be a coincidence, as similar power 
levels are also seen at larger periods and outside the stellar line at this epoch (as well as in others).  The high degree of intrinsic variability in the \hei\ line of AU~Mic, coupled to the uneven 
sampling of the observations, may indeed generate spurious signals showing up at random periods in some subsets.

\section{Summary and discussion}
\label{sec:dis}

In this paper, we analysed an extended data set of 382 unpolarized and circularly-polarized spectra of AU~Mic collected with SPIRou over a total timespan of 2041~d from early 2019 to late 2024, 
and including in particular 157 new observations recorded in 2023 and 2024 that add to the 225 already outlined and analysed in the previous study \citep{Donati23}.  We carried 
out a similar analysis of the large-scale and small-scale magnetic field of AU~Mic, of its activity and of its multi-planet system, reassessing in particular the masses of the two transiting ones 
(b and c) and the potential existence of the candidate outer planet (e) suggested in \citet{Donati23} but challenged in \citet{Mallorquin24}.  From these SPIRou spectra, we obtained time series 
of LSD Stokes $I$ and $V$ profiles, of \Bl\ values (derived from LSD signatures), of <$B$> estimates (inferred from the broadening of magnetically sensitive lines in SPIRou spectra with ZeeTurbo), 
and of RVs and $dT$ measurements (both computed with \texttt{LBL}, see Sec.~\ref{sec:obs}).  

We first find that the evolution of \Bl\ and <$B$> is similar to that previously described, with a rotation period now refined down to a precision better than 0.002~d, a factor of about two 
better than the uncertainty in our previous estimate.  We find that \Bl\ showed another minimum in the amplitude of the rotational modulation in 2023, but did not yet regain the maximum amplitude 
it reached in 2020, nor did it change its average sign (\Bl\ being on average positive over all seasons so far).  Similarly, <$B$> steadily decreased (from 2.81 to 2.64~kG) over most of the observing 
period and only started to increase in the last season, whereas the rotational modulation did not yet reach the low amplitude featured in 2020.  Besides, we confirm the 
previous finding that neither \Bl\ nor |\Bl| correlate with <$B$>.  On the contrary, <$B$> is strongly anti-correlated with $dT$, with $dT$ decreasing at a rate of 76~K/kG when <$B$> increases 
(as expected from magnetic regions being cooler than the quiet photosphere).  It also suggests that, if unspotted, AU~Mic would have a photospheric temperature of $\simeq$3875~K and that the 
average spot coverage at the stellar surface was about 34\% at the time of the monitoring \citep[assuming a typical spot-to-photosphere temperature contrast
of 620~K for AU~Mic,][]{Berdyugina05}, with rotational modulation and seasonal changes inducing periodic fluctuations of $\pm$1$-$2\%.   

Analyzing the rotational modulation of the Stokes $IV$ LSD profiles of AU~Mic with ZDI over the 11 subsets we defined (each covering up to 120~d), we find that the large-scale topology of 
AU~Mic is mainly poloidal with a quadratic average of $\simeq$1~kG. Primarily it consists of a 1.1$-$1.4~kG dipole tilted at 10$-$20\degr\ to the rotation axis (see Table~\ref{tab:mag}), 
containing 70~percent of the reconstructed magnetic energy.  The small-scale surface field <$B_I$> we derive with our self-consistent ZDI modeling, based on the assumption that <$B_I$> locally scales 
up with the large-scale field <$B_V$> (at a rate of $f_I/f_V=4.5$, see Sec.~\ref{sec:zdi}), reaches up to 5~kG at the pole.  
Its average over the visible hemisphere <$B_s$> varies from 2.5 to 3.2~kG, in agreement with the <$B$> estimates with ZeeTurbo, both regarding the long-term evolution and the amplitude of the 
rotational modulation (see Secs.~\ref{sec:bls} and \ref{sec:zdi}).  The inferred dipole strength is about twice that derived in the previous study, reflecting mostly a tighter ZDI fit to the Stokes $I$ 
LSD profiles, whereas the dipole intensity varies like <$B$>, i.e., decreasing from 2019 to a minimum in late 2022 then increasing until the end of the observations.  
The derived magnetic properties of AU~Mic and their variations with time are consistent with what has been reported on magnetic fields of active M dwarfs up to now \citep{Morin08b,Kochukhov21,Reiners22}.  

{\emr If the magnetic field of AU~Mic follows a periodic cycle like that of the Sun, the duration of this cycle must thus be significantly longer than our 6~yr monitoring, suggesting that 
the 5-yr activity cycle putatively identified by \citet{Ibanez19} may be no more than a hiccup in a longer trend.  The field evolution of AU~Mic is clearly different from and more complex than 
that of less active M dwarfs for which evidence of Sun-like magnetic cycles were reported, with global polarity switches of their poloidal field \citep{Donati23b, Lehmann24}.  
The magnetic fluctuations of AU~Mic rather resemble those of the other young (more massive) planet-hosting star V1298~Tau, whose large-scale field was shown to evolve even faster and exhibiting 
neither periodicity nor polarity switches, at least on a timescale of several years \citep{Finociety23b}.  Other reports from spectropolarimetric data covering up to 15~yr \citep[e.g.,][]{Bellotti24} similarly 
suggest that rapidly rotating M dwarfs host large-scale magnetic fields that, if periodic, have cycles longer than a decade, or are intrinsically non periodic.  } 

AU~Mic and its large-scale 
magnetic topology are conveniently oriented for Earth-based observers to detect and monitor radio emission from the stellar magnetosphere, in particular the reported highly circularly polarized and 
rotationally modulated bursts of coherent beamed emission generated by the electron cyclotron maser instability \citep{Bloot24} likely taking place in the auroral rings of the magnetic poles \citep{Hallinan15}.  
The dipole field strength we infer is consistent with estimates derived from the frequencies at which radio bursts are observed \citep[yielding 0.39 to 1.1~kG,][]{Bloot24}, implying an emission region located 
up to $\simeq$0.5~\rstar\ above the surface if generated at the fundamental cyclotron frequency, or up to $\simeq$0.9~\rstar\ if produced at the first harmonic.  

The activity of AU~Mic estimated from the \hei\ and \pab\ proxies features a dominant level of stochastic variability on top of a weak amount of rotational modulation, with 2D periodograms 
of either spectral lines in the individual subsets of the observations rarely showing a clear signal at the rotation period.  

\begin{figure}
 \centerline{\includegraphics[width=\linewidth,bb=30 10 800 530]{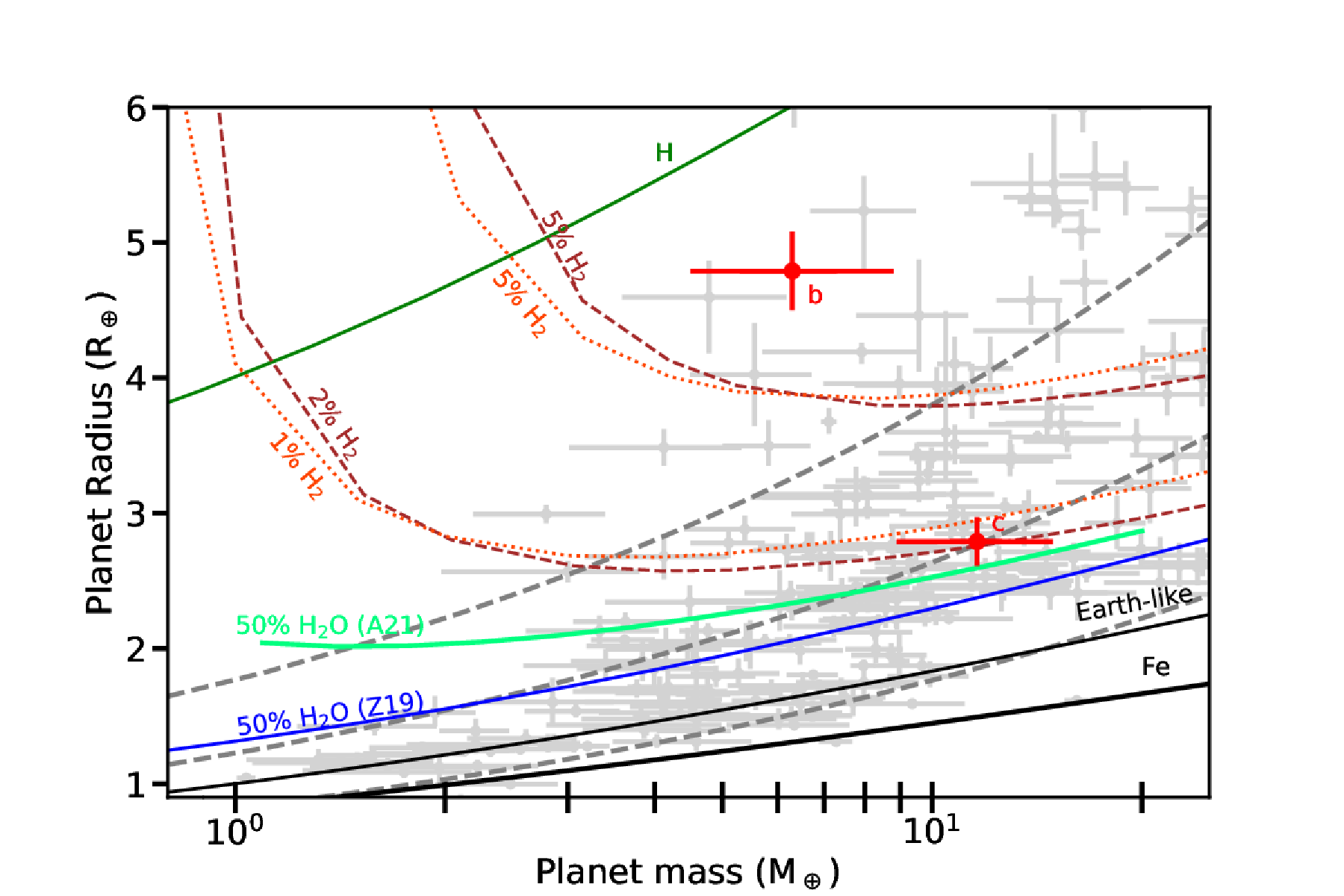}}
\caption[]{Mass-radius diagram for exoplanets with mass and radius known with a relative precision better than 30\% (grey points).  AU~Mic b and c are shown with red circles and corresponding error bars.  
Theoretical models of \citet{Zeng19} for various inner planet structures / compositions are depicted with black (100\% iron and Earth-like) and blue (50\% H$_2$O envelope) full lines, whereas those of 
\citet{Aguichine21} for the latter structures and including the effect of irradiation are shown with a full green line.  Models with a 1\%, 2\% and 5\% H$_2$ atmosphere with either an Earth-like (brown dashes) or a 50\% water 
(orange dashes) interior are also shown.  The dashed grey lines depict iso-densities of 1, 3 and 10~\gpcc\ (from top to bottom). }
\label{fig:mrp}
\end{figure}

We also exploited the SPIRou data to further constrain the masses of the planets in the AU~Mic system, the two transiting planets b and c, the candidate planet d \citep[as the potential cause of the 
reported TTVs of b and c,][]{Wittrock23,Boldog25} and the outer candidate planet e the existence of which we proposed in the previous study \citep{Donati23}.  For the first two, we inferred from the analysis RV signatures 
of semi-amplitude $K_b=2.8^{+1.1}_{-0.8}$~\ms\ and $K_c=3.9^{+1.1}_{-0.9}$~\ms, respectively corresponding to detection levels of 3.5 and 4.3$\sigma$ and to masses of $M_b=6.3^{+2.5}_{-1.8}$~\me\ and $M_c=11.6^{+3.3}_{-2.7}$~\me.  
This is smaller but consistent with the most recent estimates of \citet{Mallorquin24}.  We find no evidence than either planets have non circular orbits, with 
respective error bars on their eccentricities equal to 0.02 and 0.12 for b and c.  Whereas the orbit of b was shown to be well aligned with the equatorial plane of the star \citep{Palle20}, c may be on a 
misaligned orbit \citep{Yu25};  this would require c, presumably on an aligned orbit when still migrating within the inner disc, to have switched to another orbit at some point while maintaining 
circularity (e.g., through a resonance with an outer planet) and remaining dynamically stable \citep[unlikely for severe misalignements,][]{Yu25}.  

Using recent radius estimates for both planets ($R_b=4.79\pm0.29$~\re\ and $R_c=2.79\pm0.18$~\re\ where \re\ is the Earth's 
radius, \citealt{Mallorquin24}, in agreement with the latest measurements from \citealt{Boldog25}), we find respective densities of $0.32^{+0.13}_{-0.10}$~\gpcc\ and $2.9^{+1.1}_{-0.8}$~\gpcc, i.e., 
about 10$\times$ larger for c than for b.  Whereas c falls in the middle of the main exoplanet sequence in a radius vs mass diagram (with densities in the range 1$-$10~\gpcc), b stands clearly above 
this sequence (see Fig.~\ref{fig:mrp}), suggesting that it has not completed its contraction process yet, as a likely result of its higher equilibrium temperature \citep[possibly boosted by induction 
heating,][]{Kislyakova18} and its lower mass.  Alternatively, b may have a different composition than c, as reported for other systems \citep[e.g.,][]{Barat24}, possibly reflecting a different formation 
process if the orbit of c is indeed misaligned.  
Regarding density, b is very similar to the youngest known close-in planet recently discovered around the classical T~Tauri star IRAS~04125+2902 \citep[][]{Barber24}, for which the upper limit on the bulk 
density is even lower (0.23~\gpcc, Donati et al., submitted).  Being inflated and expected to lose mass at a 10$\times$ larger rate than c \citep{Mallorquin24}, b is an especially interesting target for 
characterizing the structure, chemistry and dynamics of young planet atmospheres \citep[e.g.,][]{Barat24}.  Attempts at detecting and estimating atmospheric escape from b through transit spectroscopy have 
however only been able to derive upper limits in the 1083~nm \hei\ triplet \citep[][]{Hirano20,Allart23,Masson24} or a variable detection in Ly$\alpha$ \citep{Rockcliffe23}, stochastic activity from the 
host star limiting the precision at which transit signatures can be characterized.  

In addition, we confirmed that the RV signature of candidate planet e discussed in the previous analysis is still present in the extended data set with a high enough confidence rate to be considered as a 
reliable detection ($\Delta \log \mathcal{L}_M=11.5$), albeit with a significantly lower semi-amplitude of $K_e=5.9^{+1.5}_{-1.2}$~\ms\ yielding a mass of $21.1^{+5.4}_{-4.3}$~\me.  The derived orbital 
period is equal to $P_e=33.11\pm0.06$~d, corresponding to a distance of 0.17~au from the host star, with the signal at $P_e$ emerging clearly in the 2D stacked periodogram of the filtered RVs as more data 
points are progressively included in the analysis.  The RV data provide no evidence that the orbit of e departs from being circular (with an error bar of 0.15 on the eccentricity), a likely condition for it 
to be dynamically stable on Gyr timescales \citep{Donati23} assuming all planets are coplanar.  Since no transit of e has yet been reported in the literature, we suspect that e does not 
transit, which implies the axis of its orbital plane to be inclined with the line of sight by no more than 88.7\degr, a value consistent with the orbital plane inclination of b and c \citep{Mallorquin24}.  If c 
is misaligned \citep{Yu25}, it may have become so under the effect of e when both fell in a 2:1 resonance in the latest phases of their migration process.  If candidate planet d proposed 
to explain the TTVs of b and c is indeed present in the system, we derive 90\% and 99\% confidence upper limits on $M_d$ respectively equal to 4.9 and 7.0~\me\ (assuming $P_d=12.73596$~d and a transit BJD 
of 2458340.55781), consistent with the estimate of \citet[][$1.0\pm0.5$~\me]{Wittrock23} or the much lower one of \citet[][$0.1$~\me]{Boldog25}.  

Given the intensity of the large-scale field (and in particular of the dipole component) we infer, the Alfven volume (in which the magnetic pressure dominates over the dynamic pressure) associated with 
AU~Mic's wind likely extends further than initially estimated \citep{Kavanagh21}, by as much as 30~percent assuming a similar topology but possibly less for a more axisymmetric case \citep{Alvarado22}.  
Following \citet{Kavanagh21}, we find that it may thus include the orbit of b for a stellar wind with a mass-loss rate up to 1000~\msw\ (where \msw\ is the mass loss rate of the Sun, i.e., $2\times10^{-14}$~\mspy) 
and that of c for a mass-loss rate up to $\simeq$250~\msw.  In this context, one can expect potential star-planet interactions to occur between AU~Mic and its innermost planet b, likely c and d as well, and 
possibly even e if the mass-loss rate is weaker than $\simeq$100~\msw.  This may be what triggers activity signatures in the 1083~nm \hei\ triplet showing up at times at the orbital period of b (see, e.g., 
right panel of Fig.~\ref{fig:eml}), similar to previous claims involving the 587~nm \hei\ line in optical of AU~Mic at the same epoch \citep{Klein22}.  It is however not clear why maximum \hei\ flux would 
occur as much as 2.4~d after the transit, nor why the \hei\ periodogram should peak at the orbital period rather than at the synodic period (of 11.4~d) if what we see corresponds to emission at the stellar 
surface from the footpoints of field lines connecting the star to planet b.  Moreover, if the reported \hei\ emission were indeed caused by star-planet interactions, one would expect to see it more or less 
all the time given the relatively stable large-scale magnetic configuration of AU~Mic.  We thus conclude that the modulated activity signal detected in the \hei\ line at the orbital period of b is more likely 
to reflect the combined effect of intrinsic variability and irregular sampling.  

Given its youth, its multi-planet system and its proximity to the Earth, AU~Mic is an obvious choice for spectroscopic, spectropolarimetric and photometric observations on a long-term monitoring basis, to 
further document the orbital and atmospheric properties of its close-in planets, characterize the magnetic field and activity of the host star, investigate potential interactions taking place between them, 
and assess the impact of massive coronal mass ejections from the host stars on the planets \citep{Alvarado22}.  We therefore advocate the stellar and exoplanet communities to continue dedicating observing 
time to the study of this key object from both ground based and space born facilities in the coming years.

\section*{Data availability}  Data exploited in this paper were collected within the SLS (up to mid 2022) and the SPICE LPs, as well as with additional PI and DDT programs.  
The SLS, DDT and pre-2023 PI data are already publicly available from the Canadian Astronomy Data Center (\url{https://www.cadc-ccda.hia-iha.nrc-cnrc.gc.ca}), whereas those from the SPICE LP 
and the 2024 PI data will be so by the end of 2025.

\begin{acknowledgements}
{\emn We thank an anonymous referee for valuable comments that improved the manuscript.} 
This study is based on data obtained at the CFHT, operated by the CNRC (Canada), INSU/CNRS (France) and the University of Hawaii. 
The authors wish to recognise and acknowledge the very significant cultural role and reverence that the summit of Maunakea has always had 
within the indigenous Hawaiian community.  
This work also benefited from the SIMBAD CDS database at \url{http://simbad.u-strasbg.fr/simbad} and the ADS system at \url{https://ui.adsabs.harvard.edu}.
\end{acknowledgements}

\bibliographystyle{aa}
\bibliography{aumic2} 
\clearpage 

\begin{appendix}

\section{Observation log}
\label{sec:appA}
Tables~\ref{tab:log} and \ref{tab:log2} provide the observation log for the \texttt{Libre ESpRIT} and \texttt{APERO} spectra respectively, and the measurements derived from them at each epoch.  

\longtab[1]{
\begin{longtable}{cccccc}
\caption{Observation log of \texttt{Libre ESpRIT} spectra. For each visit, we list the barycentric Julian date BJD, rotation cycle $c$ and phase $\phi$, the total
observing time t$_{\rm exp}$, the peak SNR in the spectrum, the noise level in the LSD Stokes $V$ profile and the estimated \Bl\ with error bars.} \label{tab:log} \\
\hline
BJD        & c / $\phi$ & t$_{\rm exp}$ & SNR & $\sigma_V$            & \Bl\  \\
(2459000+) &            &   (s)         &     & ($10^{-4} I_c$)       & (G)   \\
\hline
\endfirsthead
\caption{continued.}\\ 
\hline
BJD        & c / $\phi$ & t$_{\rm exp}$ & SNR & $\sigma_V$            & \Bl\  \\
(2459000+) &            &   (s)         &     & ($10^{-4} I_c$)       & (G)   \\
\hline
\endhead 
\hline   
\endfoot 
-399.8747695 & -83 / 0.721 & 378.9 & 427 & 1.61 & -69.3$\pm$8.0 \\
-350.8732586 & -73 / 0.804 & 763.4 & 752 & 1.25 & -62.7$\pm$6.3 \\
-349.8781098 & -72 / 0.009 & 741.1 & 753 & 1.17 & -27.0$\pm$5.7 \\
-348.9179834 & -72 / 0.206 & 735.5 & 753 & 1.39 & 116.9$\pm$6.9 \\
-348.0685108 & -72 / 0.381 & 490.3 & 407 & 1.82 & 76.6$\pm$9.3 \\
-348.0607403 & -72 / 0.383 & 490.3 & 451 & 1.67 & 66.8$\pm$8.6 \\
-348.0539511 & -72 / 0.384 & 490.3 & 511 & 1.40 & 76.8$\pm$7.1 \\
-348.0466565 & -72 / 0.385 & 490.3 & 532 & 1.37 & 64.7$\pm$7.0 \\
\end{longtable}
}

\longtab[2]{
\begin{longtable}{ccccc}
\caption{Observation log of \texttt{APERO} spectra. For each visit, we list the barycentric Julian date BJD, rotation cycle $c$
and phase $\phi$, the derived RV, small-scale field <$B$> and temperature change $dT$ (with error bars).} \label{tab:log2} \\ 
\hline
BJD        & c / $\phi$ & RV    & <$B$> & $dT$ \\
(2459000+) &            & (\ms) &  (kG) & (K)  \\ 
\hline
\endfirsthead
\caption{continued.}\\ 
\hline
BJD        & c / $\phi$ & RV    & <$B$> & $dT$ \\
(2459000+) &            & (\ms) &  (kG) & (K)  \\ 
\hline
\endhead 
\hline   
\endfoot 
-399.874709 & -83 / 0.721 & 1.4$\pm$3.7 & & -16.6$\pm$2.0 \\
-350.873275 & -73 / 0.804 & -64.6$\pm$3.3 & 2.94$\pm$0.03 & -12.6$\pm$1.8 \\
-349.873131 & -72 / 0.010 & -22.6$\pm$2.5 & 2.66$\pm$0.04 & 10.6$\pm$1.4 \\
-348.921192 & -72 / 0.206 & 31.3$\pm$2.8 & 2.75$\pm$0.04 & 0.2$\pm$1.5 \\
-347.966093 & -72 / 0.402 & 16.3$\pm$0.6 & 2.77$\pm$0.03 & -7.6$\pm$0.3 \\
-345.978216 & -72 / 0.811 & -48.3$\pm$2.2 & 2.95$\pm$0.04 & -15.1$\pm$1.2 \\
-255.178025 & -53 / 0.494 & 52.0$\pm$2.9 & 2.87$\pm$0.04 & -7.0$\pm$1.6 \\
-249.244988 & -52 / 0.715 & -6.8$\pm$3.5 & 3.08$\pm$0.04 & -17.9$\pm$1.9 \\
\end{longtable}
}

\section{Magnetic field and temperature variations of AU~Mic: additional material}
\label{sec:appA0}

\begin{figure*}[ht!]
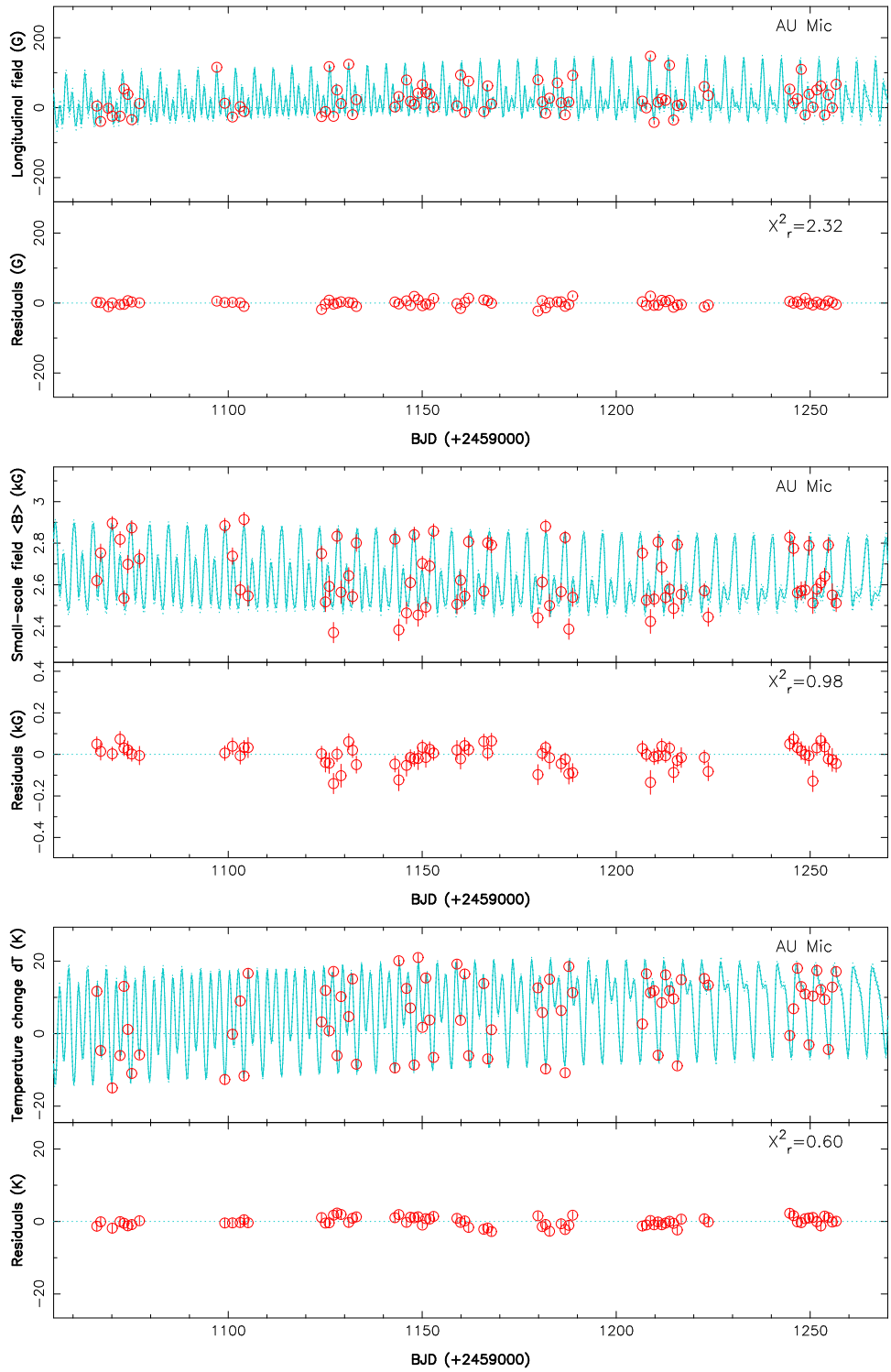

\centerline{\includegraphics[scale=0.49,angle=-90]{fignew/aumic2-gpb23.ps}\vspace{2mm}}
\centerline{\includegraphics[scale=0.49,angle=-90]{fignew/aumic2-gpbf23.ps}\vspace{2mm}}
\centerline{\includegraphics[scale=0.49,angle=-90]{fignew/aumic2-gpt23.ps}}
\caption[]{\emr Same as Fig.~\ref{fig:gpb}, zooming on the 2023 data. }
\label{fig:gpb2}
\end{figure*}

\begin{figure*}[ht!]
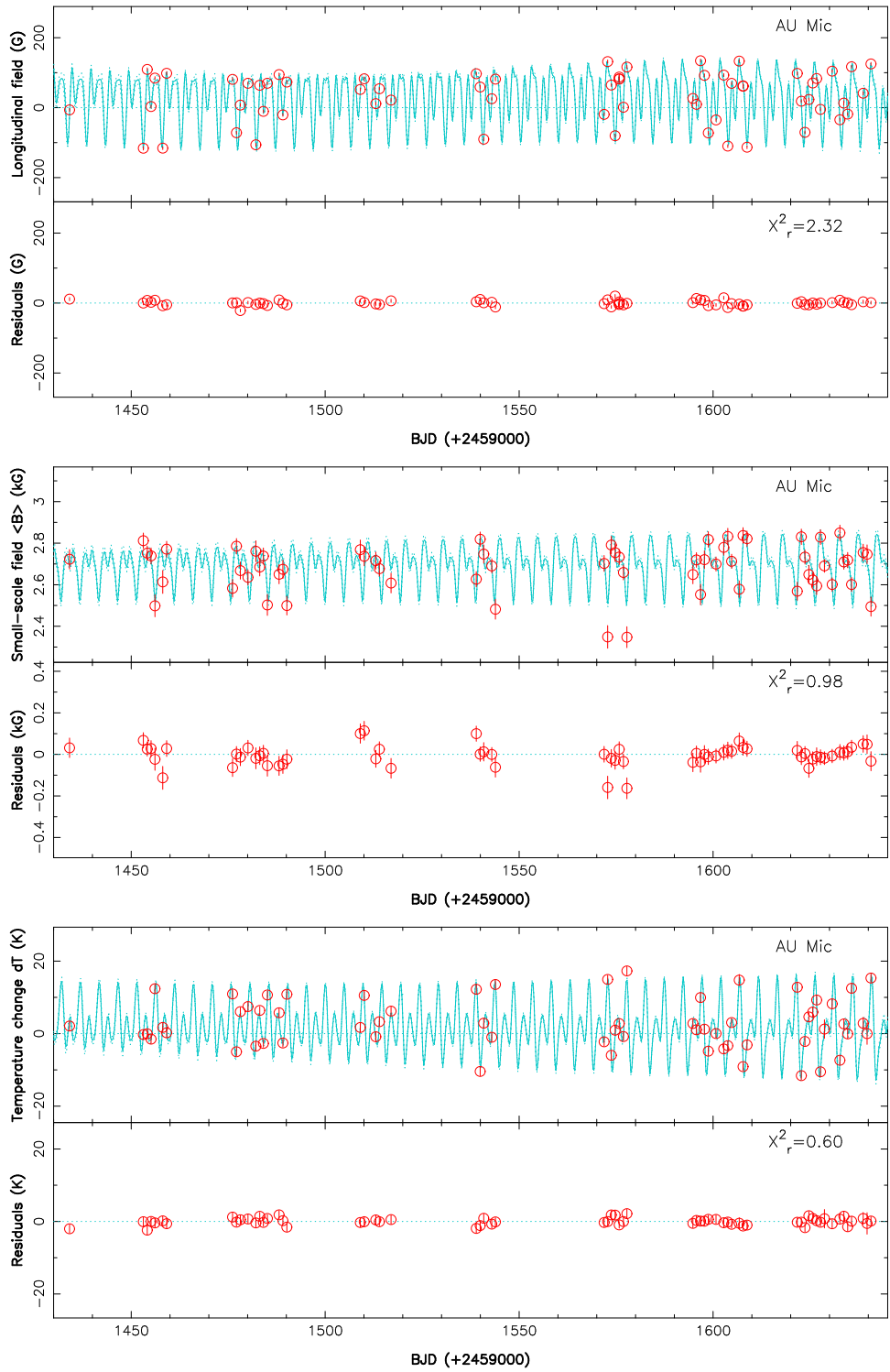

\centerline{\includegraphics[scale=0.49,angle=-90]{fignew/aumic2-gpb24.ps}\vspace{2mm}}
\centerline{\includegraphics[scale=0.49,angle=-90]{fignew/aumic2-gpbf24.ps}\vspace{2mm}}
\centerline{\includegraphics[scale=0.49,angle=-90]{fignew/aumic2-gpt24.ps}}
\caption[]{\emr Same as Fig.~\ref{fig:gpb}, zooming on the 2024 data. }
\label{fig:gpb3}
\end{figure*}

{\emr Figures~\ref{fig:gpb2} and \ref{fig:gpb3} show the \Bl, <$B$> and $dT$ curves and corresponding GPR fits, respectively zooming on the 2023 and 2024 data}, whereas Table~\ref{tab:gpr}
details the results of the GPR fit to the overall \Bl, <$B$> and $dT$ data.  

\begin{table}[ht!]
\caption{Results of the MCMC modeling of the \Bl\ (top rows), <$B$> (middle rows) and $dT$ (bottom rows) curves of AU~Mic} 
\centering
\resizebox{\linewidth}{!}{
\begin{tabular}{cccc}
\hline
Parameter   & Name & value & Prior   \\
\hline
\Bl &&&  \\ 
GP amplitude (G)     & $\theta_1$  & $84\pm7$         & mod Jeffreys ($\sigma_{\Bl}$) \\
Rec.\ period (d)     & $\theta_2$  & $4.8591\pm0.0019$  & Gaussian (4.86, 0.1) \\
Evol.\ timescale (d) & $\theta_3$  & $101\pm9$        & log Gaussian ($\log$ 100, $\log$ 2) \\
Smoothing            & $\theta_4$  & $0.38\pm0.03$    & Uniform  (0, 3) \\
White noise (G)      & $\theta_5$  & $7.9\pm0.7$      & mod Jeffreys ($\sigma_{\Bl}$) \\
Rms (G)              &             & 7.2              & \\ 
$\chisqr$            &             & 2.3              & including photon noise only \\ 
\hline
<$B$> &&&  \\ 
GP amplitude (kG)    & $\theta_1$  & $0.135\pm0.016$  & mod Jeffreys ($\sigma_{<B>}$) \\
Rec.\ period (d)     & $\theta_2$  & $4.8633\pm0.0016$  & Gaussian (4.86, 0.1) \\
Evol.\ timescale (d) & $\theta_3$  & $208\pm21$       & log Gaussian ($\log$ 200, $\log$ 2) \\
Smoothing            & $\theta_4$  & $0.43\pm0.04$    & Uniform  (0, 3) \\
White noise (kG)     & $\theta_5$  & $0.01\pm0.01$    & mod Jeffreys ($\sigma_{<B>}$) \\
Rms (kG)             &             & 0.038            & \\ 
$\chisqr$            &             & 0.98             & including photon noise only \\ 
\hline
$dT$ &&&  \\ 
GP amplitude (K)     & $\theta_1$  & $10.5\pm1.2$     & mod Jeffreys ($\sigma_{dT}$) \\
Rec.\ period (d)     & $\theta_2$  & $4.8611\pm0.0015$  & Gaussian (4.86, 0.1) \\
Evol.\ timescale (d) & $\theta_3$  & $167\pm11$       & log Gaussian ($\log$ 150, $\log$ 2) \\
Smoothing            & $\theta_4$  & $0.46\pm0.03$    & Uniform  (0, 3) \\
White noise (K)      & $\theta_5$  & $0.2\pm0.1$      & mod Jeffreys ($\sigma_{dT}$) \\
Rms (K)              &             & 0.95             & \\ 
$\chisqr$            &             & 0.60             & including photon noise only \\ 
\hline
\end{tabular}}
\tablefoot{\emr For each hyper parameter, 
we list the fitted value along with the corresponding error bar, as well as the assumed prior.  The knee of the modified Jeffreys prior is set to
the median error bars of the \Bl, <$B$> and $dT$ estimates (i.e., 4.7~G, 0.038~kG and 1.3~K respectively).  For the recurrence period
$\theta_2$, using a uniform prior yields the same result.  For the evolution timescale $\theta_3$, the log Gaussian prior is set to a value ranging from
100 to 200~d (within a factor of 2), determined from preliminary runs. } 
\label{tab:gpr}
\end{table}

\section{ZDI modeling of AU~Mic: additional material}
\label{sec:appA1}

\begin{figure*}[ht!]
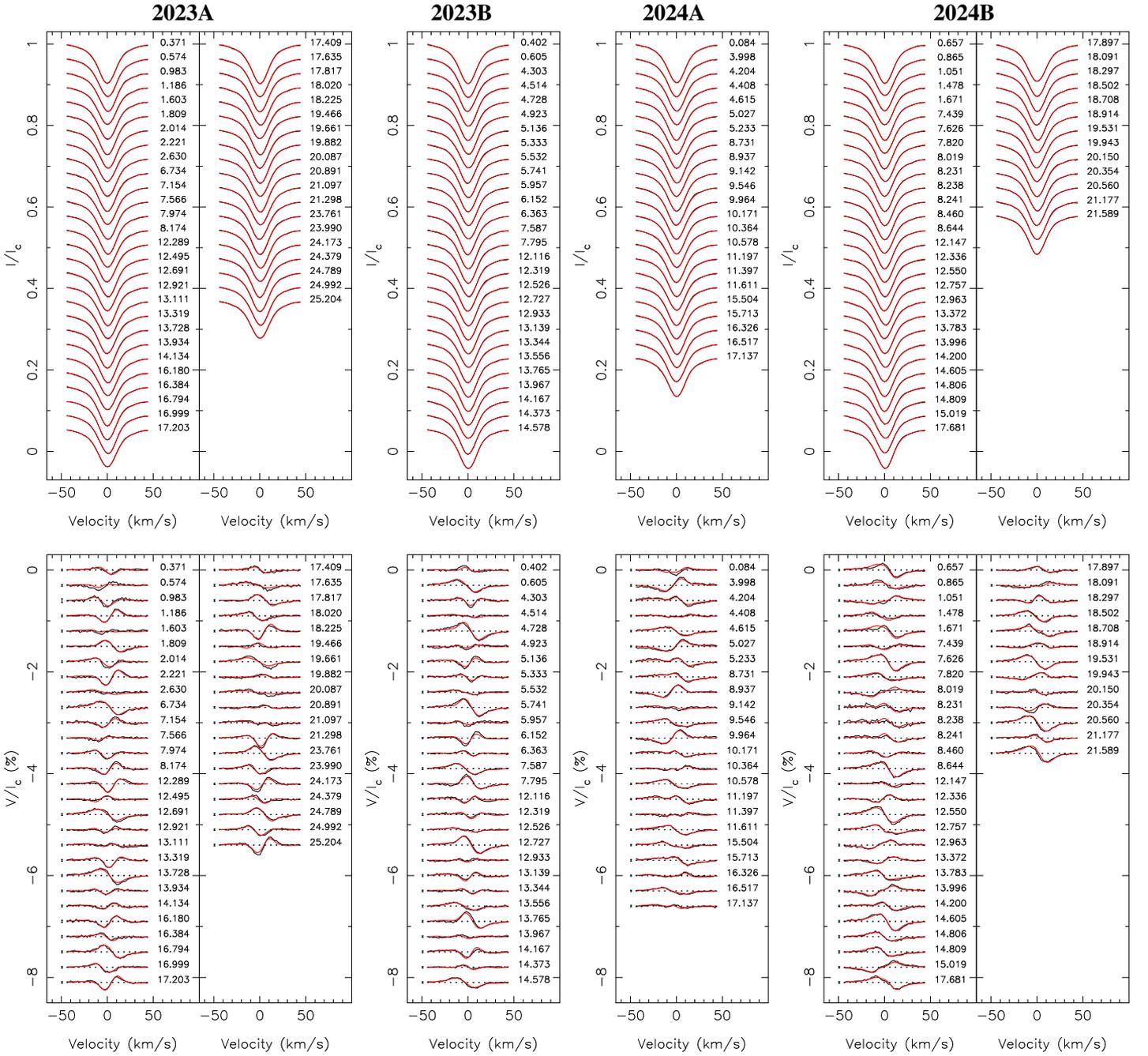

\flushleft{\large\bf \hspace{2.3cm}2023A\hspace{3.6cm}2023B\hspace{2.5cm}2024A\hspace{3.8cm}2024B\vspace{-2mm}}
\centerline{\includegraphics[scale=0.25,angle=-90]{fignew/aumic2-fiti23a.ps}\hspace{1mm}
            \includegraphics[scale=0.25,angle=-90]{fignew/aumic2-fiti23b.ps}\hspace{1mm}
            \includegraphics[scale=0.25,angle=-90]{fignew/aumic2-fiti24a.ps}\hspace{1mm}
            \includegraphics[scale=0.25,angle=-90]{fignew/aumic2-fiti24b.ps}}
\flushleft{\vspace{1mm}} 
\centerline{\includegraphics[scale=0.25,angle=-90]{fignew/aumic2-fitv23a.ps}\hspace{1mm}
            \includegraphics[scale=0.25,angle=-90]{fignew/aumic2-fitv23b.ps}\hspace{1mm}
            \includegraphics[scale=0.25,angle=-90]{fignew/aumic2-fitv24a.ps}\hspace{1mm}
            \includegraphics[scale=0.25,angle=-90]{fignew/aumic2-fitv24b.ps}}
\caption[]{Observed (thick black line) and modelled (thin red line) LSD Stokes $V$ profiles of the photospheric lines of AU~Mic, for subsets 2023A, 
2023B, 2024A and 2024B (from left to right).  The ZDI modeling of these profiles is described in 
Sec.~\ref{sec:zdi}.  Rotation cycles (counting from 219, 244, 295 and 316 for 2023A, 2023B, 2024A and 2024B respectively, see Table~\ref{tab:log}) 
are indicated to the right of all profiles, whereas $\pm$1$\sigma$ error bars are shown to the left of LSD Stokes $V$ profiles.  } 
\label{fig:fit}
\end{figure*}

\begin{figure*}[ht!]
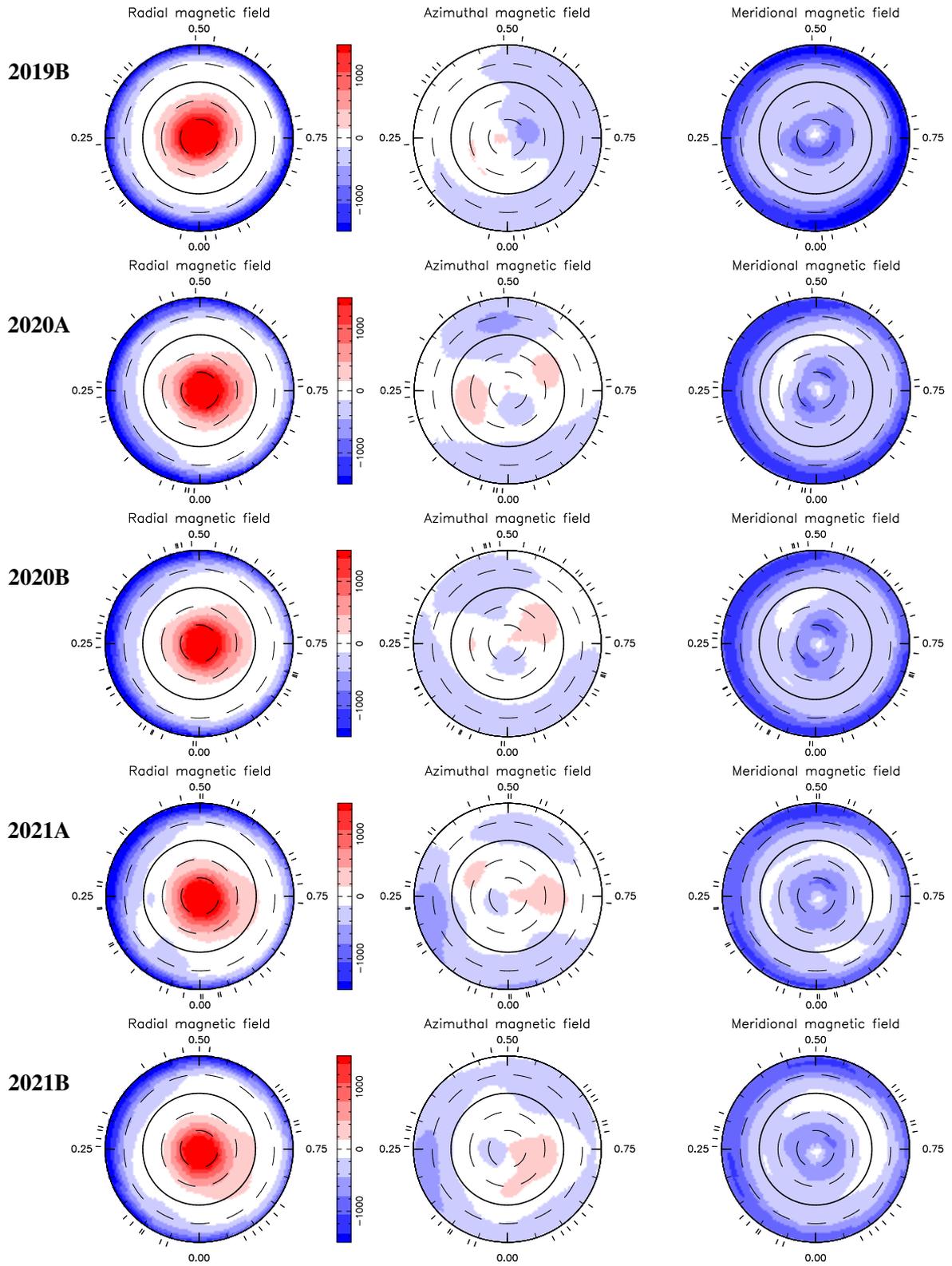

\centerline{\large\bf 2019B\raisebox{0.3\totalheight}{\includegraphics[scale=0.40,angle=-90]{fignew/aumic2-map19b.ps}}\vspace{1mm}}
\centerline{\large\bf 2020A\raisebox{0.3\totalheight}{\includegraphics[scale=0.40,angle=-90]{fignew/aumic2-map20a.ps}}\vspace{1mm}}
\centerline{\large\bf 2020B\raisebox{0.3\totalheight}{\includegraphics[scale=0.40,angle=-90]{fignew/aumic2-map20b.ps}}\vspace{1mm}}
\centerline{\large\bf 2021A\raisebox{0.3\totalheight}{\includegraphics[scale=0.40,angle=-90]{fignew/aumic2-map21a.ps}}\vspace{1mm}} 
\centerline{\large\bf 2021B\raisebox{0.3\totalheight}{\includegraphics[scale=0.40,angle=-90]{fignew/aumic2-map21b.ps}}}
\caption[]{Same as Fig.~\ref{fig:map} for seasons 2019B, 2020A, 2020B, 2021A and 2021B.} 
\label{fig:map2}
\end{figure*}

\begin{figure*}[ht!]
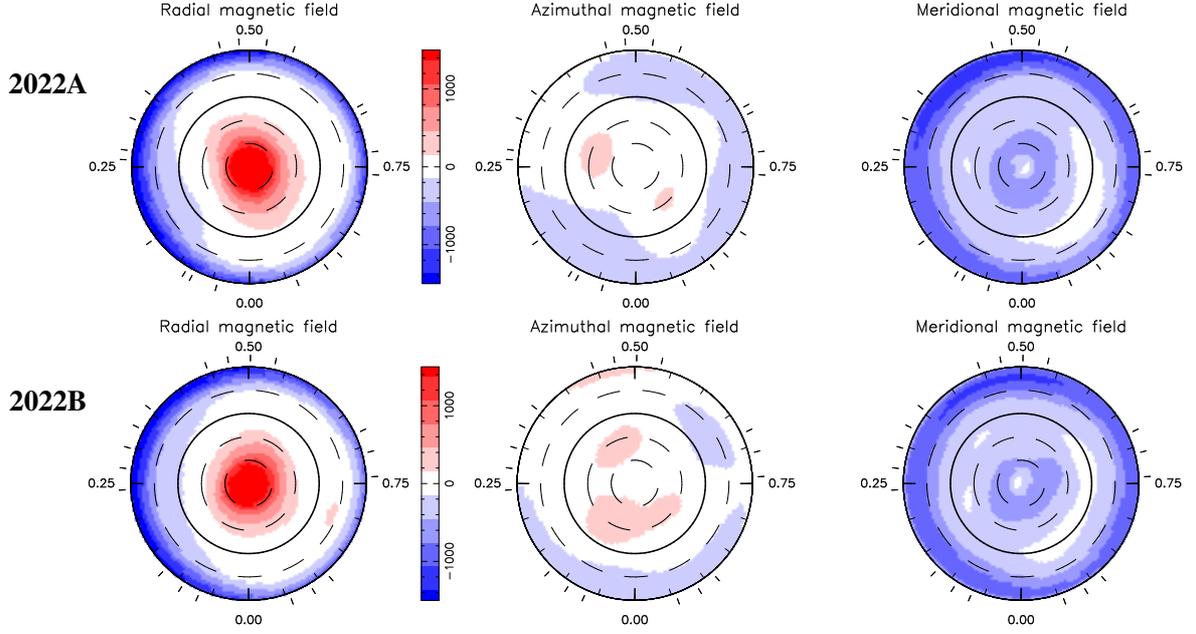

\centerline{\large\bf 2022A\raisebox{0.3\totalheight}{\includegraphics[scale=0.40,angle=-90]{fignew/aumic2-map22a.ps}}\vspace{1mm}}
\centerline{\large\bf 2022B\raisebox{0.3\totalheight}{\includegraphics[scale=0.40,angle=-90]{fignew/aumic2-map22b.ps}}}
\caption[]{Same as Fig.~\ref{fig:map} for seasons 2022A and 2022B.} 
\label{fig:map3}
\end{figure*}

Figure~\ref{fig:fit} shows the observed and reconstructed LSD Stokes $I$ and $V$ profiles of AU~Mic for seasons 2023A to 2024B, 
whereas Figs.~\ref{fig:map2} and \ref{fig:map3} show the reconstructed ZDI maps of AU~Mic for seasons 2019B to 2022B.  
Table~\ref{tab:mag} recaps the main characteristics of the magnetic topologies reconstructed with ZDI for all subsets.  

\begin{table}[ht!]
\caption{Properties of the large-scale and small-scale magnetic field of AU~Mic derived with ZDI, for the 11 data subsets} 
\centering
\resizebox{\linewidth}{!}{
\begin{tabular}{ccccccc}
\hline
         & \multicolumn{6}{c}{Stokes $IV$ analysis}                      \\
         & \multicolumn{6}{c}{($f_I=0.9$, $f_V=0.2$, $\vD=3.5$~\kms)}    \\
\hline
Season   & <$B_V$> & <$B_I$> & <$B_s$> & \Bd  & tilt / phase & pol/axi   \\
         &  (kG)   & (kG)    & (kG)   & (kG) & (\degr / ) & (percent) \\
\hline
2019B    & 1.15   & 5.2   & 3.18 / 0.46 & 1.38 & 10 / 0.52 & 97 / 97 \\
2020A    & 1.04   & 4.7   & 2.91 / 0.15 & 1.31 & 16 / 0.74 & 98 / 93 \\
2020B    & 1.03   & 4.6   & 2.79 / 0.14 & 1.29 & 16 / 0.79 & 98 / 93 \\
2021A    & 1.00   & 4.5   & 2.70 / 0.41 & 1.29 & 18 / 0.82 & 97 / 92 \\
2021B    & 0.93   & 4.2   & 2.52 / 0.50 & 1.18 & 17 / 0.84 & 97 / 93 \\
2022A    & 0.92   & 4.1   & 2.52 / 0.50 & 1.15 & 14 / 0.78 & 97 / 95 \\
2022B    & 0.86   & 3.9   & 2.46 / 0.51 & 1.08 & 16 / 0.80 & 99 / 95 \\
2023A    & 0.98   & 4.4   & 2.66 / 0.33 & 1.18 &  9 / 0.54 & 98 / 98 \\
2023B    & 0.95   & 4.3   & 2.50 / 0.29 & 1.17 & 11 / 0.57 & 98 / 98 \\
2024A    & 0.97   & 4.4   & 2.63 / 0.25 & 1.20 & 11 / 0.61 & 98 / 98 \\
2024B    & 1.04   & 4.7   & 2.78 / 0.33 & 1.25 & 11 / 0.61 & 98 / 97 \\
\hline
\end{tabular}}
\tablefoot{\emr Columns 2 and 3 respectively list the large-scale and small-scale fields, once quadratically                                  
averaged over the whole stellar surface.  Column 4 gives the time-averaged and full-amplitude variations of the small scale field
integrated over the visible hemisphere <$B_s$> (as if measured from Zeeman broadening of lines profiles), whereas columns 5 to 7 respectively list the polar
strengths of the dipole component \Bd, the tilt of the dipole component to the rotation axis and the phase towards which it is tilted, and finally the amount
of magnetic energy reconstructed in the poloidal component of the field and in the axisymmetric modes of this component.  Error bars on field values and
percentages are typically equal to 10~percent, whereas dipole tilts are accurate to typically 5\degr. }
\label{tab:mag}
\end{table}

\section{RVs of AU~Mic: additional material}
\label{sec:appA2}

\begin{figure*}[ht!]
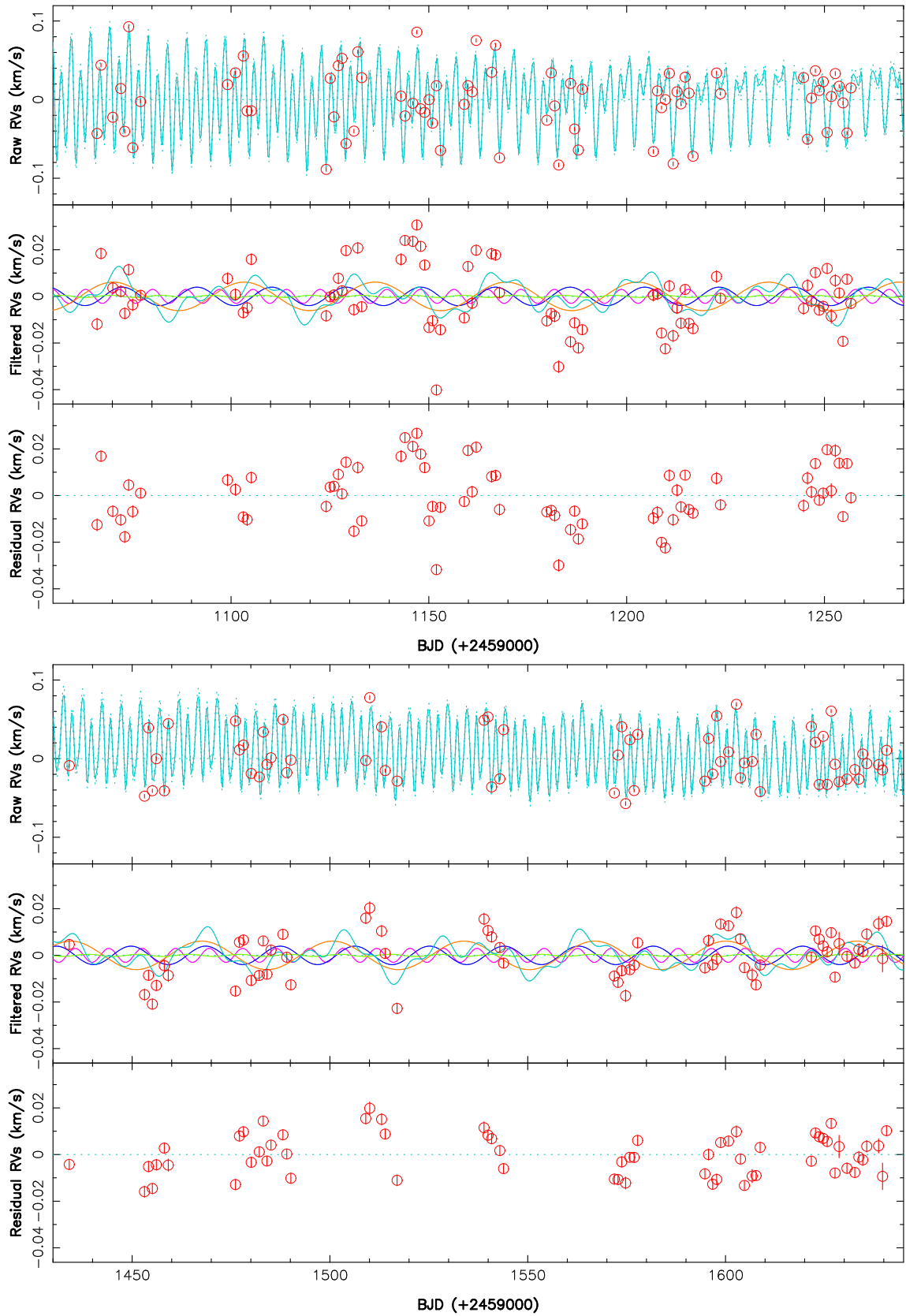

\centerline{\includegraphics[scale=0.58,angle=-90]{fignew/aumic2-rv4-23.ps}\vspace{1mm}}
\centerline{\includegraphics[scale=0.58,angle=-90]{fignew/aumic2-rv4-24.ps}}
\caption[]{\emr Same as Fig.~\ref{fig:rvr}, zooming on the 2023 (top) and 2024 (bottom) data. } 
\label{fig:rvr2}
\end{figure*}

{\emr Figure~\ref{fig:rvr2} shows the RV curve of AU~Mic and the corresponding 4-planet and GPR fits, zooming on the 2023 and 2024 data.} 
{\emn Table~\ref{tab:pla2} details the results of the multi-dimensional GPR fit to the $dT$ and RV data mentioned in Sec.~\ref{sec:act}. } 

\begin{table*}[ht!]
\caption{\emn Same as Fig.~\ref{tab:pla} for a multi-dimensional GPR fit to both $dT$ and RVs.} 
\centering
\resizebox{0.8\textwidth}{!}{
\begin{tabular}{cccccc}
\hline
Parameter          & No planet                 & b+c                       & b+c+e                     & b+c+e+d                   &   Prior \\ 
\hline
$\theta_1$ (K)     & $10.9\pm1.2$         & $10.9\pm1.2$         & $11.0\pm1.2$         & $11.1\pm1.2$         & mod Jeffreys ($\sigma_{\rm dT}$) \\ 
$\theta_2$ (d)     & $4.8601\pm0.0021$    & $4.8596\pm0.0023$    & $4.8595\pm0.0023$    & $4.8595\pm0.0022$    & Gaussian (4.86, 0.1) \\ 
$\theta_3$ (d)     & $132^{+25}_{-21}$    & $121^{+21}_{-18}$    & $118^{+19}_{-16}$    & $119^{+18}_{-16}$    & log Gaussian ($\log$ 120, $\log$ 1.5) \\ 
$\theta_4$         & $0.52\pm0.04$        & $0.52\pm0.04$        & $0.52\pm0.04$        & $0.53\pm0.04$        & Uniform  (0, 3) \\ 
$\theta_5$ (K)     & $0.2\pm0.2$          & $0.2\pm0.2$          & $0.2\pm0.2$          & $0.2\pm0.2$          & mod Jeffreys ($\sigma_{\rm dT}$) \\ 
$\theta_6$ (\ms)   & $0.9^{+0.9}_{-0.5}$  & $0.9^{+0.8}_{-0.4}$  & $0.9^{+0.8}_{-0.4}$  & $0.9^{+0.8}_{-0.4}$  & mod Jeffreys ($\sigma_{\rm RV}$) \\  
$\theta_7$ (\ms)   & $22.6^{+2.8}_{-2.5}$ & $22.6^{+2.7}_{-2.4}$ & $22.8^{+2.7}_{-2.4}$ & $23.1^{+2.7}_{-2.4}$ & mod Jeffreys ($\sigma_{\rm RV}$) \\  
$\theta_8$ (\ms)   & $17.9\pm0.8$         & $17.2\pm0.8$         & $16.9\pm0.8$         & $16.8\pm0.8$         & mod Jeffreys ($\sigma_{\rm RV}$) \\ 
\hline
$K_b$ (\ms)        &                      & $3.1^{+1.6}_{-1.1}$  & $3.0^{+1.6}_{-1.0}$  & $3.1^{+1.6}_{-1.1}$  & mod Jeffreys ($\sigma_{\rm RV}$) \\ 
$P_b$ (d)          &                      & 8.463446             & 8.463446             & 8.463446                  & fixed from \citet{Mallorquin24} \\ 
BJD$_b$ (2459000+) &                      & $-669.649175$        & $-669.649175$             & $-669.649175$             & fixed from \citet{Mallorquin24} \\ 
\hline
$K_c$ (\ms)        &                      & $4.5^{+1.5}_{-1.1}$  & $4.3^{+1.5}_{-1.1}$  & $4.1^{+1.6}_{-1.1}$  & mod Jeffreys ($\sigma_{\rm RV}$) \\ 
$P_c$ (d)          &                      & 18.859018            & 18.859018            & 18.859018                 & fixed from \citet{Mallorquin24} \\ 
BJD$_c$ (2459000+) &                      & $-657.776513$        & $-657.776513$        & $-657.776513$             & fixed from \citet{Mallorquin24} \\ 
\hline
$K_e$ (\ms)        &                      &                           & $5.1^{+1.7}_{-1.2}$  & $4.9^{+1.7}_{-1.3}$  & mod Jeffreys ($\sigma_{\rm RV}$) \\  
$P_e$ (d)          &                      &                           & $33.15\pm0.08$       & $33.15\pm0.10$       & Gaussian (33.1, 1.0) \\ 
BJD$_e$ (2459000+) &                      &                           & $118.6\pm1.6$        & $118.9\pm1.9$        & Gaussian (118, 8) \\ 
\hline
$K_d$ (\ms)        &                 &                           &                           & $1.3^{+1.2}_{-0.6}$  & mod Jeffreys ($\sigma_{\rm RV}$) \\  
$P_d$ (d)          &                 &                           &                           & 12.73596             & fixed from \citet{Wittrock23} \\ 
BJD$_d$ (2459000+) &                 &                           &                           & $-659.44219$         & fixed from \citet{Wittrock23} \\ 
\hline
\chisqr\ ($dT$)    & 0.55                 & 0.53                 & 0.55                 & 0.53                 &  \\ 
\chisqr\ (RVs)     & 50.5                 & 46.9                 & 44.4                 & 44.0                 &  \\ 
rms (\ms)          & 16.7                 & 16.1                 & 15.6                 & 15.6                 &  \\ 
$\log \mathcal{L}_M$ & 115.9              & 125.8                & 136.2                & 137.8                &  \\
$\log {\rm BF} = \Delta \log \mathcal{L}_M$ & $-9.9$  & 0.0      & 10.4                 & 12.0                 &  \\
\hline 
\end{tabular}}
\tablefoot{\emn For the multi-dimensional GPR fit, including three additional parameters, $\theta_1$ refers to the amplitude of the $dT$ component ($F$ term), 
$\theta_6$ and $\theta_7$ to the amplitudes of the RV component ($F$ and $F'$ terms), $\theta_5$ and $\theta_8$ to the excess jitter of the $dT$ and RV components, 
whereas the other terms keep the same meaning as in Table~\ref{tab:pla} \citep[for more information, see][]{Rajpaul15}.  We note that $\theta_6$ is close to 0, 
confirming that for AU~Mic, RVs are highly correlated with the first derivative of $dT$ (see Sec.~\ref{sec:act})}  
\label{tab:pla2}
\end{table*}

\section{Activity of AU~Mic: additional material}
\label{sec:appB}

\begin{figure*}[ht!]
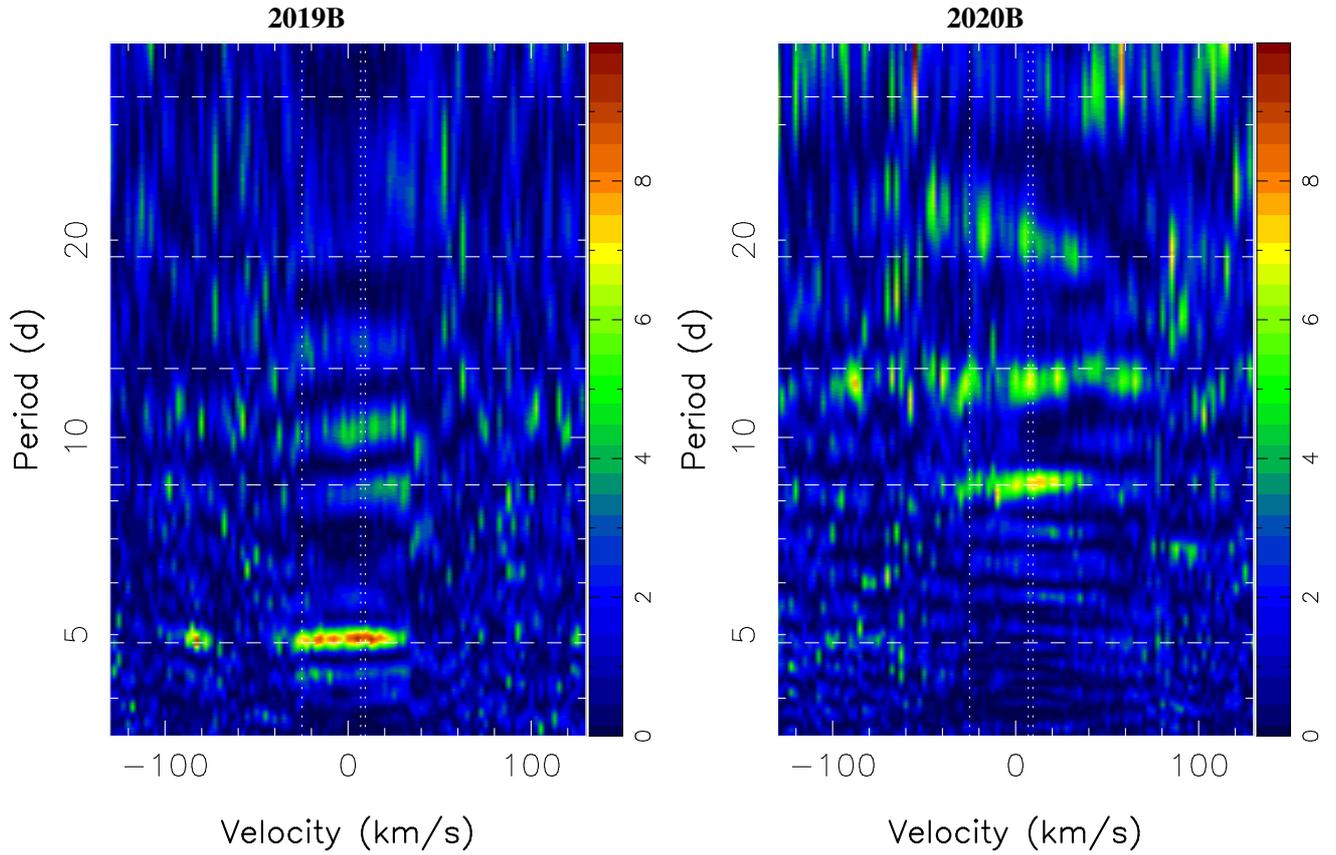

\flushleft{\bf\large \hspace{4.0cm}2019B\hspace{7.9cm}2020B\vspace{-2mm}}
\centerline{\includegraphics[scale=0.55,angle=-90]{fignew/aumic2-heiper19b.ps}\hspace{3mm}\includegraphics[scale=0.55,angle=-90]{fignew/aumic2-heiper20b.ps}}
\caption[]{2D periodograms of the 1083.3-nm \hei\ triplet residual (with respect to the median) in the stellar rest frame, for the 2019B (left) and 2020B (right) spectra of 
AU~Mic.  In both periodograms, the dashed horizontal line traces \Prot\ and the orbital periods of the transiting and candidate planets, whereas the vertical dotted lines 
depict the velocities of the three components of the \hei\ triplet.  The color scale traces the logarithmic power in the periodograms.  
We caution that only the main peaks (colored yellow to red and extending over at least several velocity bins) are likely to be significant in these plots.   } 
\label{fig:eml}
\end{figure*}

Figure~\ref{fig:eml} depicts the 2D periodograms of the 1083.3~nm \hei\ line in the spectrum of AU~Mic, in semesters 2019B and 2020B.  

\FloatBarrier 
\clearpage

\end{appendix}
\end{document}